\documentclass{article} 
\usepackage{iclr2026/iclr2026_conference,times}

\usepackage{amsmath,amsfonts,bm}

\def\eqref#1{equation~\ref{#1}}

\def\1{\bm{1}}

\DeclareMathAlphabet{\mathsfit}{\encodingdefault}{\sfdefault}{m}{sl}
\SetMathAlphabet{\mathsfit}{bold}{\encodingdefault}{\sfdefault}{bx}{n}

\usepackage{hyperref}
\usepackage{url}
\usepackage{graphicx}
\usepackage{wrapfig}
\usepackage{booktabs}
\usepackage{multirow}
\usepackage{siunitx} 
\usepackage{graphicx}   
\usepackage{booktabs}   
\usepackage{caption}    
\usepackage{subcaption} 

\title{CLoD-GS: Continuous Level-of-Detail via 3D Gaussian Splatting}

\author{
    \makebox[\textwidth][c]{
    \begin{tabular}{c} 
    Zhigang Cheng\textsuperscript{1}\thanks{Work done during internship at AMAP.} \quad
    Mingchao Sun\textsuperscript{2} \quad
    Yu Liu\textsuperscript{2} \quad
    Zengye Ge\textsuperscript{2} \\
    Luyang Tang\textsuperscript{2} \quad
    Mu Xu\textsuperscript{2}\thanks{Co-corresponding authors.} \quad
    Yangyan Li\textsuperscript{2}\footnotemark[2] \quad
    Peng Pan\textsuperscript{1}\footnotemark[2]
    \\[2.5ex] 
    \textsuperscript{1}Tsinghua University \qquad
    \textsuperscript{2}AMAP \qquad
    \\[1.5ex] 
    \texttt{xumu.xm@alibaba-inc.com, yangyan.lee@gmail.com, pengpan@tsinghua.edu.cn}
    \end{tabular}}
}

\iclrfinalcopy 

\begin{document}

\maketitle
\begin{abstract}

Level of Detail (LoD) is a fundamental technique in real-time computer graphics for managing the rendering costs of complex scenes while preserving visual fidelity. Traditionally, LoD is implemented using discrete levels (DLoD), where multiple, distinct versions of a model are swapped out at different distances. However, this long-standing paradigm suffers from two major drawbacks: it requires significant storage for multiple model copies and causes jarring visual ``popping" artifacts during transitions, degrading the user experience. We argue that the explicit, primitive-based nature of the emerging 3D Gaussian Splatting (3DGS) technique enables a more ideal paradigm: Continuous LoD (CLoD). A CLoD approach facilitates smooth and seamless quality scaling within a single unified model, thereby circumventing the core problems of DLOD. To this end, we introduce CLoD-GS, a framework that integrates a continuous LoD mechanism directly into a 3DGS representation. Our method introduces a learnable distance-dependent decay parameter for each Gaussian primitive that dynamically adjusts its opacity based on viewpoint proximity. This allows for the progressive and smooth filtering of less significant primitives, effectively creating a continuous spectrum of detail within one model. To train this model to be robust across all distances, we introduce a virtual distance scaling mechanism with point count regularization. Our approach not only eliminates the storage overhead and visual artifacts of discrete methods but also reduces the primitive count and memory footprint of the final model. Extensive experiments demonstrate that CLoD-GS achieves smooth, quality-scalable rendering from a single model, delivering high-fidelity results across a wide range of performance targets.
\end{abstract}

\section{Introduction}

The pursuit of photorealism in real-time computer graphics is characterized by a fundamental tension between ever-increasing scene complexity and the finite computational budget of rendering hardware. To maintain interactive frame rates, systems must intelligently manage the number of primitives rendered per frame, a challenge first articulated decades ago \citep{funkhouser1993adaptive}. This constrained optimization problem—generating the best possible image within a fixed time budget—has driven the development of Level of Detail (LoD) techniques, which adaptively reduce an object's complexity based on its perceptual importance to the viewer \citep{luebke2002level}.

The most established paradigm for LoD is Discrete Level of Detail (DLoD). In this approach, artists or automated algorithms create multiple, distinct versions of a model at varying complexities~\citep{clark1976hierarchical}. At runtime, the system selects the appropriate version based on metrics like distance or screen-space projection size. While computationally efficient, DLoD suffers from two critical drawbacks. First, storing multiple copies of every asset leads to a significant memory overhead, limiting scene scale and variety. Second, the instantaneous swap between discrete models causes jarring visual ``popping" artifacts, degrading the user experience~\citep{giegl2007unpopping}.

The recent advent of 3D Gaussian Splatting (3DGS) has revolutionized novel view synthesis, achieving state-of-the-art visual quality at real-time rendering speeds \citep{kerbl20233d}. By representing scenes as a collection of explicit 3D Gaussian primitives, 3DGS leverages a highly optimized rasterization pipeline. However, this paradigm does not escape the fundamental constraint of rendering cost; performance still scales with the number of primitives, making LoD a necessity for complex scenes. Given that 3DGS is a primitive-based representation, a straightforward approach would be to apply the traditional DLoD paradigm by creating multiple, discrete sets of Gaussians at varying levels of detail \citep{kulhanek2025lodge}. While feasible, this strategy inevitably reintroduces the classic DLoD drawbacks: a significant storage overhead for maintaining multiple Gaussian clouds and the visually disruptive ``popping" artifacts during transitions.

We argue that a more ideal paradigm could be achieved with the unique characteristics of the 3DGS: Continuous Level of Detail (CLoD). The profound suitability of 3DGS for a CLoD framework stems from its fundamental representational properties, which distinguish it from traditional discrete primitives like meshes or point clouds. First, each Gaussian is not a discrete point but a continuous volumetric entity—a probability distribution with a ``soft" footprint. This makes modulating its contribution to the scene (e.g., via its opacity) an intrinsically smooth and continuous operation. In contrast, simplifying a mesh requires discrete topological changes like edge collapses\citep{hoppe1996progressive}. Second, each primitive is defined by a set of continuous parameters that can be finely controlled. This allows for per-primitive filtering rather than the abrupt removal of entire geometric elements. Finally, the entire representation is end-to-end differentiable. This feature allows the LoD mechanism itself to be learned. We can introduce new learnable parameters that control simplification and optimize them directly within the primary training process.

To this end, we introduce CLoD-GS, a framework integrating a continuous LoD mechanism directly into the 3DGS representation. Our key contribution is to augment each Gaussian primitive with an additional learnable parameter: a distance-dependent decay factor. This parameter dynamically modulates the primitive's opacity based on its proximity to the viewpoint, allowing for the smooth filtering of less significant details, effectively creating a continuous spectrum of detail within a single, unified model.
To train a single model that is robust across the entire LoD spectrum, we introduce a novel virtual distance scaling training strategy. This involves rendering from virtually scaled distances to activate the LoD mechanism, with a point count regularization loss that explicitly encourages the model to learn a more compact representation for distant views. 
Our framework eliminates the storage overhead and popping artifacts inherent to discrete methods while simultaneously reducing the final model's primitive count. Experiments demonstrate that CLoD-GS achieves high-fidelity results across a wide range of performance targets, paving the way for more scalable and visually coherent real-time neural rendering applications.

\section{Related Work}

\textbf{LoD.} The earliest and most common approach, DLoD, involves pre-generating multiple versions of a mesh at different resolutions \citep{clark1976hierarchical}. At runtime, the system selects an appropriate model based on heuristics like distance or screen-space size \citep{funkhouser1993adaptive}. The creation of these simplified meshes spurred a rich field of research in polygonal simplification algorithms \citep{schroeder1992decimation, garland1997surface}. DLoD's reliance on multiple asset copies leads to high storage costs and visually jarring ``popping" artifacts during transitions \citep{luebke2002level}. Techniques were developed to mitigate popping, but often at the cost of increased rendering complexity or other visual artifacts \citep{giegl2007unpopping}. To address the shortcomings of DLoD, CLoD techniques were developed. The foundational work in this area is Progressive Meshes by \citet{hoppe1996progressive}, which represents a mesh as a coarse base model plus a sequence of vertex split operations that can incrementally refine it. This concept was extended to View-Dependent Simplification, where the LoD can vary locally across a single object's surface based on viewing parameters \citep{hoppe1997view, luebke2000view}. While CLoD successfully eliminates storage overhead and popping, it shifts complexity from the asset pipeline to the runtime algorithm, often incurring significant CPU overhead to traverse the hierarchical data structures and generate the appropriate mesh each frame \citep{lindstrom2002real}. 

\textbf{Neural Scene Representations.} The field of novel view synthesis was revolutionized by Neural Radiance Fields (NeRF), mapping 5D coordinates (position and viewing direction) to volumetric density and color \citep{mildenhall2020nerf}. In the context of NeRF, the LoD problem manifests primarily as aliasing when viewing scenes at different scales. This was addressed by \citet{barron2021mipnerf} in Mip-NeRF. Other works like Strata-NeRF have explored imposing discrete structures onto implicit fields, representing a form of discrete LoD for NeRFs \citep{dhiman2023strata}. The development of methods like Instant-NGP accelerated NeRF training times, making these implicit methods more practical \citep{mueller2022instant}. To bridge the gap between the quality of implicit methods and the speed of traditional rasterization, explicit neural representations were developed. Early works like Plenoxels \citep{fridovich2022plenoxels} and TensoRF \citep{chen2022tensorf} discretized the scene into explicit representations, enabling much faster training and rendering. The most significant breakthrough in this area has been 3DGS \citep{kerbl20233d}. The return to an explicit, primitive-based representation makes the vast body of knowledge on geometric LoD directly relevant once again, creating a fertile ground for new frameworks like ours.

\textbf{LoD for 3D Gaussian Splatting.} Several works have adapted the DLoD philosophy to 3DGS by creating hierarchical representations. A notable example is LODGE \citep{kulhanek2025lodge}, which creates multiple, discrete sets of Gaussians by iteratively applying smoothing filters and importance-based pruning. Hierarchical-3DGS \citep{hierarchicalgaussians24} focuses on designing hierarchical data structures and streaming systems for massive-scale scenes. Similarly, Octree-GS \citep{ren2024octree} organizes the scene into an octree, assigning ``anchor Gaussians" to different levels of the hierarchy. While these methods provide structured control over detail, they inherit the complexities of traditional DLoD. They rely on rigid, explicit data structures that add algorithmic and memory overhead, and their management of discrete levels can cause popping artifacts. The objectives of these methods are thus different from ours. Consequently, our method can complement these large-scale DLoD systems by offering finer-grained, internal quality control for each loaded data chunk or level of the hierarchy. In the CLoD field, \citet{milef2025learning} propose a method that achieves continuous LoD through a separate, subsequent training phase. After an initial 3DGS model is trained, they perform an additional optimization stage where random subsets of splats are used for rendering. 

\textbf{3DGS Model Compression.} It is important to distinguish dynamic LoD techniques from static model compression. While LoD is a runtime optimization for performance scaling, compression is a pre-process for reducing storage and transmission costs. Recent work in 3DGS compression includes pruning-based methods like LightGaussian, which permanently remove redundant Gaussians from a trained model \citep{fan2023lightgaussian}, and quantization-based methods like Compact3D and CompGS, which reduce the precision of Gaussian attributes using techniques like vector quantization \citep{lee2024compact, navaneet2024compgs}. A sophisticated pruning approach is presented by MaskGaussian \citep{wu2024maskedgaussian}, which models each Gaussian as a probabilistic entity with a learnable ``probability of existence" to guide a more robust pruning process. Fundamentally, these are static compression techniques designed to produce a single, smaller model. 

\begin{figure}[h]
\centering 
\includegraphics[width=0.88\textwidth]{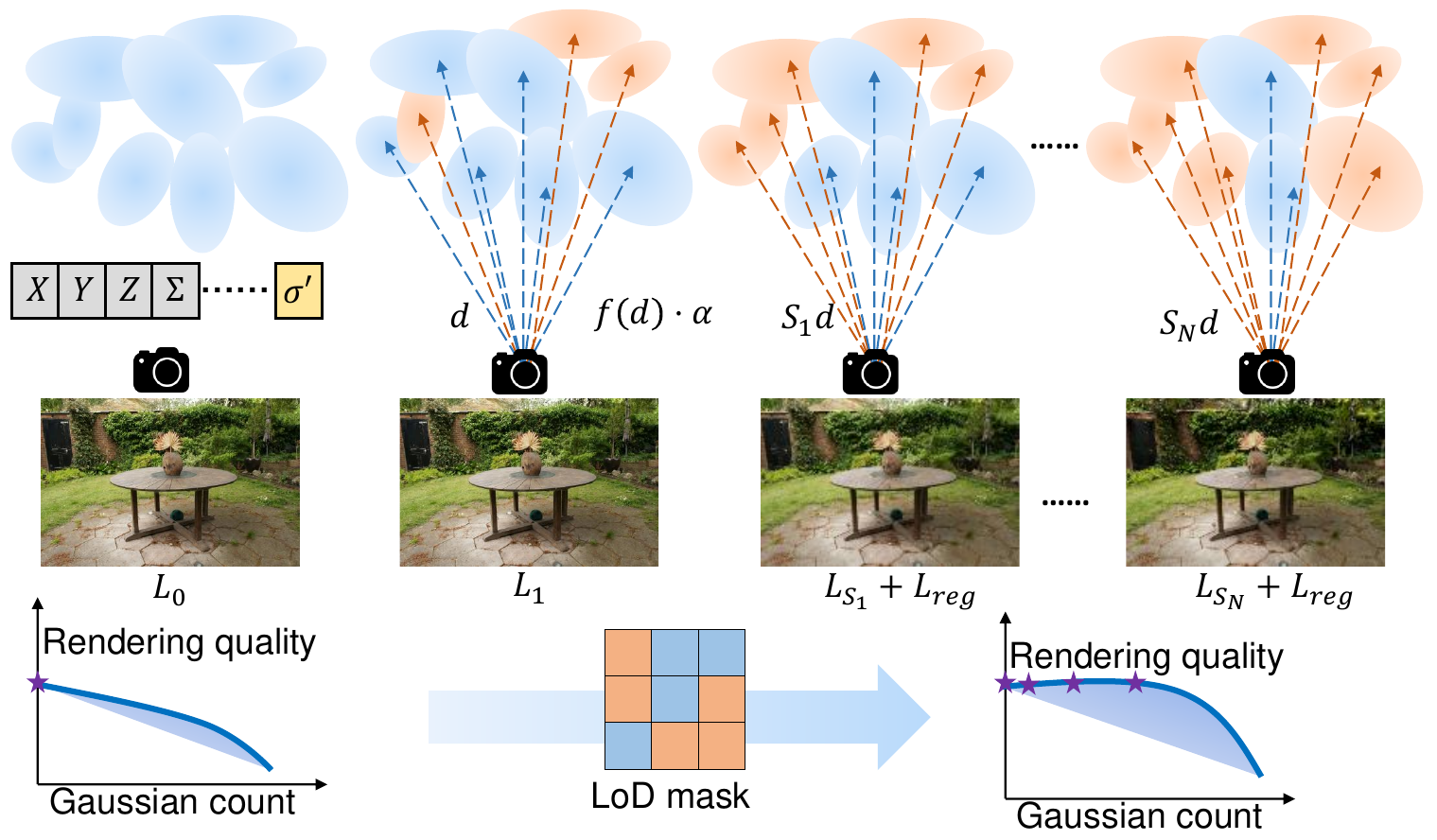} 
\caption{Framework of the proposed methodology.} 
\label{fig_frame} 
\end{figure}
\section{Methodology}
Our core objective is to develop a single, unified 3DGS model that can dynamically and smoothly adjust its LoD at render time. This allows for a seamless trade-off between visual quality and performance, catering to diverse hardware capabilities and user requirements. As illustrated in Figure~\ref{fig_frame}, our framework achieves this by building upon the standard 3DGS representation with three key innovations: a novel learnable parameterization for distance-adaptive opacity, a dynamic masking mechanism driven by the viewing distance, and a specialized virtual distance scaling training strategy to ensure robustness across all detail levels.

\subsection{Preliminaries: 3D Gaussian Splatting}
The standard 3DGS framework \citep{kerbl20233d} represents a 3D scene as a collection of explicit, anisotropic 3D Gaussian primitives. Each Gaussian $i$ is defined by a set of learnable attributes: a 3D position (mean) $\mu_i$, a 3D covariance matrix $\Sigma_i$ (represented by scaling and rotation factors), a base opacity $\alpha_i$, and coefficients for Spherical Harmonics (SH) to model view-dependent color $c_i$.
Rendering is achieved through a highly optimized, differentiable rasterization process. For a given viewpoint, 3D Gaussians are projected onto the 2D image plane. The final color $C$ for a pixel is then computed by alpha-blending the projected Gaussians sorted by depth:
\begin{equation}
C = \sum_{i \in N} c_i \alpha'_i \prod_{j=1}^{i-1}(1-\alpha'_j)
\end{equation}
where $N$ is the set of Gaussians overlapping the pixel, and $\alpha'_i$ is the effective 2D opacity, which is a product of the base opacity $\alpha_i$ and the 2D Gaussian's value at the pixel center. This explicit, primitive-based, and differentiable representation serves as the foundation for our proposed LoD mechanism.

\subsection{Learnable Continuous LoD via Distance-Adaptive Opacity}
To achieve a continuous LoD, our goal is to smoothly modulate the contribution of each Gaussian primitive based on its perceptual importance, which is strongly correlated with its distance from the viewer. Instead of discretely removing primitives, which causes popping, we propose to dynamically attenuate their opacity, allowing them to fade out gracefully. This leverages the continuous nature of the Gaussian representation and integrates seamlessly into the alpha-blending pipeline.

To this end, we introduce a single additional learnable parameter for each Gaussian primitive $i$: the distance decay factor $\sigma_{d,i}$. This scalar parameter is optimized alongside the Gaussian's other attributes and learns to control how rapidly the primitive's visibility should decrease with distance.
At render time, for a given camera center $c$, we first compute the Euclidean distance $d_i = ||\mu_i - c||$ for each Gaussian. To ensure the decay effect is consistent across different scenes and camera perspectives, we normalize this distance. We define the normalized distance $d'_i = {d_i} / {\max_{j \in N_{\text{view}}} (d_j)}$,
where $N_{\text{view}}$ is the set of all Gaussians within the current view frustum. We then use this normalized distance, along with a user-controllable virtual distance scale factor $s_v$, to compute the attenuated opacity $\alpha_i''$:
\begin{equation}
\alpha_i'' = \alpha_i \cdot \exp\left(-\frac{(d'_i \cdot s_v)^2}{2 \cdot (\text{ReLU}(\sigma_{d,i}))^2 + \epsilon}\right)
\end{equation}
where $\alpha_i$ is the original learned opacity. The virtual distance scale $s_v \geq 1$ allows the user to simulate the effect of viewing the scene from farther away, thereby increasing the attenuation. The $\text{ReLU}(\cdot)$ function ensures the learned decay parameter $\sigma_{d,i}$ remains non-negative, and $\epsilon$ is a small constant for numerical stability. This formulation effectively weights the original opacity with a Gaussian falloff whose variance is learned per-primitive, a functional form widely used to model distance-dependent effects \citep{takikawa2021neural,strugar2009continuous}.

After computing the attenuated opacity $\alpha_i''$, we apply a dynamic threshold to determine which primitives are significant enough to be sent to the rasterizer. This filtering is defined by a boolean mask $M_i = (\alpha_i'' > \tau \cdot s_v)$,
where $\tau$ is a small base opacity threshold. By scaling the threshold with $s_v$, we apply a stricter culling criterion when simulating more distant views. Only Gaussians with $M_i = 1$ are rendered. This simple, per-primitive computation allows for a continuous and smooth performance-quality trade-off by adjusting the single scalar $s_v$.

\subsection{Virtual Distance Scaling Training for a Unified LoD Model}
A key challenge in creating a unified LoD model is training it to perform effectively across the entire spectrum of detail, from full fidelity to aggressive simplification. A model trained only on high-quality views would not learn a meaningful simplification behavior. To address this, we propose a virtual distance scaling training strategy that forces the model to learn a representation that is robust at all viewing scales.

During each training iteration, we randomly sample a virtual distance scale factor $s_v$ from a predefined range (e.g., $\mathcal{U}(1, 10)$), where $\mathcal{U}$ denotes a uniform distribution. This strategy compels the model to optimize its parameters not only for the ground-truth camera views (where $s_v=1$) but also for simulated distant views ($s_v > 1$).

However, we observed that relying solely on rendering from virtual distances was insufficient, as the model could converge to trivial solutions. We therefore introduce a more direct form of supervision through a primitive count regularization loss. This loss explicitly encourages the model to utilize fewer primitives for more distant views. To this end, we define a target primitive ratio, $\eta_{\text{target}}$, which is inversely proportional to the virtual distance scale:
\begin{equation}
    \eta_{\text{target}} = 1 / s_v^{1.5}.
\end{equation}
The exponent $1.5$ is an empirically determined value that controls the rate of geometric simplification. We then formulate a regularization loss, $L_{\text{reg}}$, which penalizes the model if the actual ratio of rendered primitives, $\eta_{\text{actual}}$, exceeds this target:
\begin{equation}
    L_{\text{reg}} = (s_v - 1.0)^2 \cdot \left( \text{ReLU}(\eta_{\text{actual}} - \eta_{\text{target}}) \right)^2,
\end{equation}
where $\eta_{\text{actual}} = (\sum_{i} M_i) / N_{\text{total}}$. Here, $N_{\text{total}}$ is the total number of primitives in the scene, and $M_i \in \{0, 1\}$ is a binary mask indicating whether the $i$-th primitive is rendered. The quadratic weight term $(s_v - 1.0)^2$ ensures that this penalty is applied only to simulated distant views ($s_v > 1$) and that its magnitude increases with the virtual distance.

The final training objective combines the standard 3DGS rendering loss, $L_{\text{render}}$ (a weighted sum of L1 and D-SSIM losses), with our regularization term:
\begin{equation}
    L_{\text{total}} = w_s (L_{\text{render}} + \lambda_{\text{reg}} L_{\text{reg}}).
\end{equation}
Here, the weight $w_s = (1 - 0.5 \cdot s_v / \max(s_v))^2$ is designed to apply weaker supervision to renderings from larger virtual distances. This prevents the model from over-pruning primitives at the expense of rendering quality. The hyperparameter $\lambda_{\text{reg}}$ balances the trade-off between reconstruction fidelity and the sparsity constraint. This comprehensive training strategy equips the model with the ability to learn an efficient, view-dependent representation, thereby enabling a single, robust model with controllable and continuous Level-of-Detail (LoD) capabilities.

\section{Experiments}
\label{sec:experimets}
We conduct a series of comprehensive experiments to validate the effectiveness of our CLoD-GS framework. We first detail the experimental setup, then present quantitative and qualitative comparisons against state-of-the-art methods, and finally provide in-depth ablation and robustness studies.

\subsection{Experimental Setup}

\textbf{Datasets.} Our evaluations are performed on 12 real-world scenes from three challenging public datasets: the BungeeNeRF dataset \citep{xiangli2022bungeenerf} (8 scenes), the Tanks and Temples dataset \citep{knapitsch2017tanks} (2 scenes), the Deep Blending dataset \citep{hedman2018deep} (2 scenes) and the MipNeRF360 dataset \citep{barron2022mipnerf360}. For all experiments, we follow the original 3DGS train/test split to ensure fair comparisons.

\textbf{Evaluation Metrics.} We use three standard metrics for novel view synthesis: Peak Signal-to-Noise Ratio (PSNR), Structural Similarity Index (SSIM) \citep{wang2004image}, Learned Perceptual Image Patch Similarity (LPIPS) \citep{zhang2018unreasonable} and Frames Per Second (FPS).

\textbf{Implementation Details.} Our framework is built on the official 3DGS implementation. All experiments are run on an Ubuntu server with four NVIDIA RTX 4090 GPUs. We train our models for 30,000 iterations and enable the proposed mechanism since 5000 iterations. The learning rate for our learnable distance decay factor $\sigma_{d,i}$ is 1e-2, and the weight for the regularization loss $\lambda_{\text{reg}}$ is set to 1.0. Notably, our method utilizes the same set of hyperparameters across all datasets. The scale parameter, tested at values of 1, 3, 5 and 7, defines the maximum allowable value for the $s_v$ during training. For our experiments, both 3DGS and MaskGaussian were implemented on the latest public 3DGS codebase. Unless otherwise specified, MaskGaussian uses the `beta' settings from its original paper \citep{wu2024maskedgaussian}.

\begin{wraptable}{r}{0.5\textwidth}
    \centering
    \caption{Quantitative comparison of highest-quality models. Best results are \textbf{bold}, second best are \underline{underlined}. The fifth and sixth columns indicate the number of Gaussian primitives (\#GS) and memory consumption (Mem). `$\downarrow$' indicates that lower is better.}
    \label{tab:main_results}
    \vspace{1ex}
    \resizebox{0.5\textwidth}{!}{%
    \begin{tabular}{lrrrrr}
    \toprule
    \textbf{Method} & \textbf{PSNR} $\uparrow$ & \textbf{SSIM} $\uparrow$ & \textbf{LPIPS} $\downarrow$ & \textbf{\#GS(k)} $\downarrow$ & \textbf{Mem(MB)} $\downarrow$ \\
    \midrule
    \multicolumn{6}{c}{\textit{BungeeNeRF}} \\
    \midrule
    3DGS              & 27.91          & \underline{0.917}    & \textbf{0.096}    & 6733          & 1592.48 \\
    Fast Rendering    & /              & /                    & /                 & 6733          & 1592.48 \\
    Octree-GS         & \underline{27.94} & 0.909                & 0.110             & /  & 1045.70 \\
    MaskGaussian      & 27.76          & 0.916                & 0.098             & 5298          & 1253.13 \\
    Ours (scale=1)    & \textbf{28.05} & \textbf{0.919}       & \underline{0.100} & 4185          & 1005.87 \\
    Ours (scale=3)    & 27.70          & 0.908                & 0.117             & \underline{2738}  & \underline{658.01} \\
    Ours (scale=7)    & 27.09          & 0.885                & 0.150             & \textbf{1855} & \textbf{445.72}\\
    \midrule
    \multicolumn{6}{c}{\textit{Tanks\&Temples}} \\
    \midrule
    3DGS              & 23.70          & \underline{0.853}    & \underline{0.169} & 1574          & 372.19 \\
    Fast Rendering    & 23.62          & \underline{0.853}    & 0.194             & 1574          & 372.19 \\
    Octree-GS         & \textbf{24.17} & \textbf{0.858}       & \textbf{0.161}    & /             & 383.90 \\
    MaskGaussian      & 23.56          & 0.846                & 0.180             & 1237          & 292.68 \\
    H-3DGS ($\tau$=0) & 21.71          & 0.820                & 0.200             & /             & / \\
    H-3DGS ($\tau$=3) & 21.77          & 0.821                & 0.200             & /             & / \\
    H-3DGS ($\tau$=6) & 21.76          & 0.818                & 0.206             & /             & / \\
    H-3DGS ($\tau$=15)& 21.55          & 0.800                & 0.239             & /             & / \\
    Ours (scale=1)    & 23.75          & 0.843                & 0.185             & 1159          & 278.53 \\
    Ours (scale=3)    & \underline{23.79} & 0.843                & 0.185             & \underline{984}  & \underline{236.58} \\
    Ours (scale=7)    & 23.67          & 0.839                & 0.193             & \textbf{884}  & \textbf{212.54} \\
    \midrule
    \multicolumn{6}{c}{\textit{Deep Blending}} \\
    \midrule
    3DGS              & 29.84          & 0.907                & \textbf{0.238}    & 2486          & 587.98 \\
    Fast Rendering    & 29.00          & 0.902                & 0.303             & 2486          & 587.98 \\
    Octree-GS         & 29.65          & 0.901                & 0.257             & /             & \underline{180.00} \\
    MaskGaussian      & 29.66          & 0.907                & 0.244             & 1778          & 420.41 \\
    H-3DGS ($\tau$=0) & 27.41          & 0.887                & 0.254             & /             & / \\
    H-3DGS ($\tau$=3) & 27.40          & 0.887                & 0.254             & /             & / \\
    H-3DGS ($\tau$=6) & 27.38          & 0.887                & 0.255             & /             & / \\
    H-3DGS ($\tau$=15)& 27.26          & 0.884                & 0.265             & /             & / \\
    Ours (scale=1)    & \textbf{29.93} & \textbf{0.908}       & \underline{0.239} & 1697          & 407.72 \\
    Ours (scale=3)    & \underline{29.86} & \underline{0.907}    & 0.244             & \underline{1258} & 302.27 \\
    Ours (scale=7)    & 29.64          & \underline{0.907}    & 0.251             & \textbf{662}  & \textbf{159.04} \\
    \bottomrule
    \end{tabular}
    }
\end{wraptable}

\textbf{Compared Methods.} We compare CLoD-GS against several leading methods:
 \textbf{3DGS} \citep{kerbl20233d}: The original method, serving as the high-quality, high-cost baseline.
\textbf{Fast Rendering} \citep{milef2025learning}: A state-of-the-art continuous LoD method based on learning a static importance ranking for splats.
\textbf{Octree-GS} \citep{ren2024octree} and \textbf{H-3DGS} \citep{hierarchicalgaussians24}: State-of-the-art discrete LoD methods using hierarchical structures.
\textbf{MaskGaussian} \citep{wu2024maskedgaussian}: A state-of-the-art static compression method that uses probabilistic masks for robust pruning.
For Fast Rendering and Octree-GS, we report the metrics from their respective papers. All other methods were trained locally under identical conditions for a fair comparison.

\subsection{Results and Comparisons}

\begin{wraptable}{r}{0.5\textwidth}
    \centering
    \caption{Comparison of FPS on various datasets. Best results are \textbf{bold}, second best are \underline{underlined}. Higher is better ($\uparrow$).}
    \label{tab:fps_results}
    \vspace{1ex}
    \resizebox{0.5\textwidth}{!}{%
    \begin{tabular}{lrrrr}
    \toprule
    \textbf{Method} & \textbf{BungeeNeRF} $\uparrow$ & \textbf{Tanks\&Temples} $\uparrow$ & \textbf{Deep Blending} $\uparrow$ & \textbf{Mip-NeRF 360} $\uparrow$ \\
    \midrule
    Octree-GS         & \underline{76.05} & 134.87             & 141.59             & \underline{129.45} \\
    H-3DGS ($\tau$=0) & /               & 42.30              & 36.30              & / \\
    H-3DGS ($\tau$=3) & /               & 54.05              & 46.50              & / \\
    H-3DGS ($\tau$=6) & /               & 61.54              & 47.37              & / \\
    H-3DGS ($\tau$=15)& /               & 75.82              & 51.03              & / \\
    Ours (scale=1)    & 57.34             & 169.89             & 128.68             & 109.42 \\
    Ours (scale=3)    & 69.74             & \underline{176.74} & \underline{145.83} & 125.33 \\
    Ours (scale=7)    & \textbf{87.88}    & \textbf{199.12}    & \textbf{187.48}    & \textbf{140.58} \\
    \bottomrule
    \end{tabular}
    }
\end{wraptable}

Table~\ref{tab:main_results} and Table~\ref{tab:fps_results} summarize the performance of all methods at their highest quality settings. Our CLoD-GS consistently achieves higher rendering quality and speed, often surpassing the original 3DGS in PSNR and SSIM while using significantly fewer Gaussians (e.g., a 38\% reduction on BungeeNeRF). This demonstrates that our regularization strategy inherently produces a more compact and efficient representation. While Octree-GS achieves the highest compression ratios, it introduces considerable rendering overhead for smaller scenes and is not natively supported by existing renderers. The superior performance of our method is particularly pronounced on datasets with large depth variations and significant focal length changes, such as BungeeNeRF. This highlights the effectiveness of our designed LoD mechanism in handling complex, multi-scale scenes.
\begin{figure}[h!]
\centering
\includegraphics[width=\textwidth]{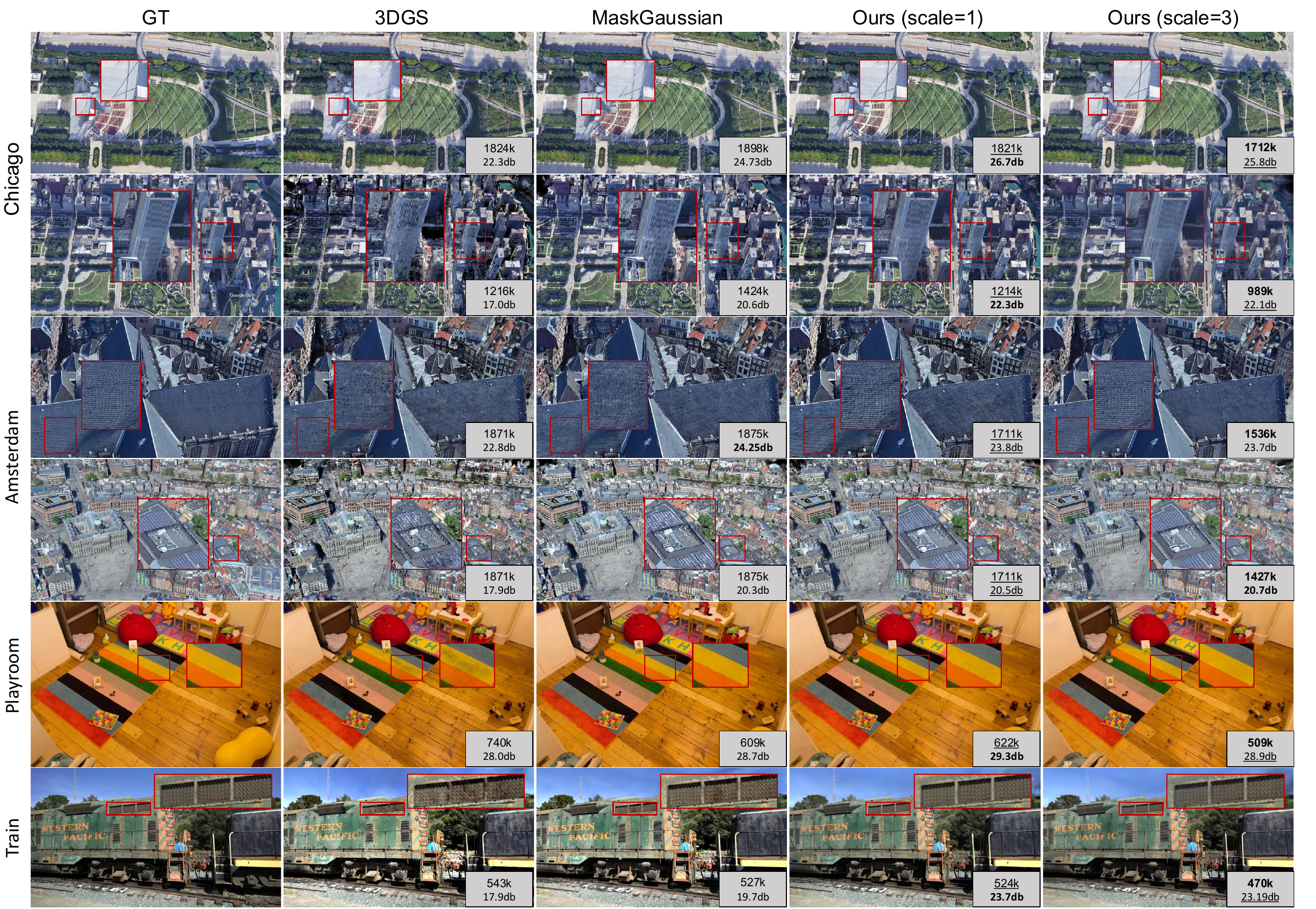}
\caption{Visual comparison at similar primitive counts. The number of Gaussians used and the corresponding PSNR are annotated in the bottom-right corner of each image.}
\label{fig:qualitative_comp}
\end{figure}

\begin{figure}[h!]
\centering
\includegraphics[width=\textwidth]{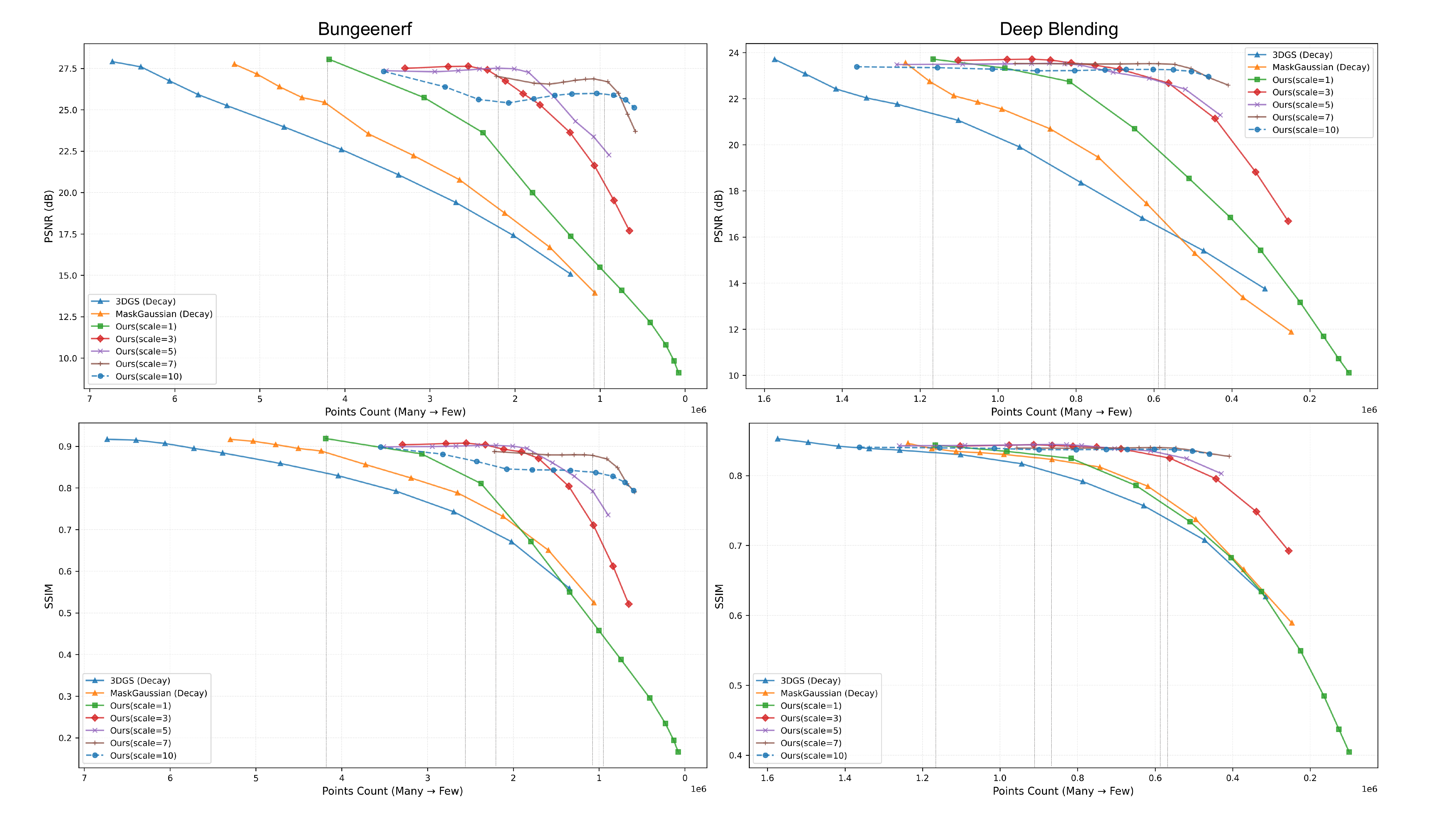}
\caption{Quality vs. primitive count on the BungeeNeRF and Deep Blending datasets. The dashed lines indicate the maximum virtual scale during training. Our method with varying virtual distance scale ranges ($s_v$) shows a more graceful quality degradation.}
\label{fig:quality_curves}
\end{figure}

\begin{figure}[h!]
\centering
\includegraphics[width=\textwidth]{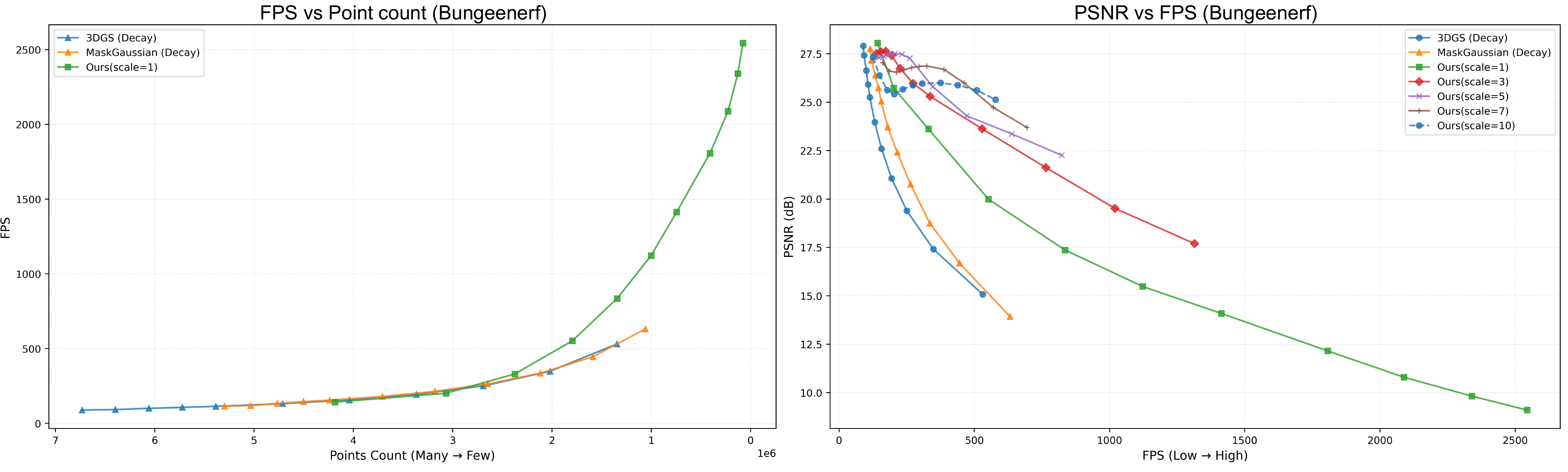}
\caption{Relationship between FPS, the number of Gaussians, and rendering quality. As shown, rendering speed exhibits a strong negative correlation with the number of Gaussians. Reducing the number of Gaussians boosts FPS for all methods, with our method showing a more significant increase.}
\label{fig:fps_curves}
\end{figure}

As shown in Figure~\ref{fig:qualitative_comp}, our method produces clearer results at similar primitive counts, particularly in areas with repetitive textures or complex lighting, while often using fewer Gaussians. 
Figure~\ref{fig:quality_curves} and Figure~\ref{fig:fps_curves} illustrate the trade-off between rendering quality and speed as a function of the number of Gaussians. For a fair comparison, the points for 3DGS and MaskGaussian are also selected by applying our opacity attenuation formula (Equation 3) to their trained models and selecting the primitives with the highest resulting opacities. The curves for CLoD-GS show that increasing the virtual distance scale range during training (e.g., from $s_v \in [1, 3]$ to $s_v \in [1, 5]$) produces models that are more robust to simplification. Notably, this improved low-detail performance is achieved while maintaining nearly identical peak quality, confirming our method's ability to create a single, versatile model that performs well across the entire quality-performance spectrum. Ultimately, when pruning Gaussians to enhance FPS, our method consistently achieves higher rendering quality.

\begin{wraptable}{r}{0.4\textwidth} 
    \centering
    \caption{Ablation study on our key training components. The full model outperforms all ablated versions.}
    \label{tab:ablation}
    \vspace{1ex} 
    \resizebox{0.4\textwidth}{!}{
    \begin{tabular}{lrrr} 
    \toprule
    \textbf{Method} & \textbf{PSNR} $\uparrow$ & \textbf{SSIM} $\uparrow$ & \textbf{LPIPS} $\downarrow$ \\ 
    \midrule
    \multicolumn{4}{c}{\textit{BungeeNeRF}} \\ 
    \midrule
    Full Model (Ours) & \textbf{27.59} & \textbf{0.902} & \textbf{0.123} \\ 
    w/o weight ($w_s$)& 27.39         & \underline{0.894} &\underline{0.127} \\ 
    w/o loss ($L_{\text{reg}}$)&\underline{27.56}&\textbf{0.902} & \textbf{0.123} \\ 
    w/o weight \& loss& 26.71      &0.871          & 0.169      \\ 
    \midrule
    \multicolumn{4}{c}{\textit{Deep Blending}} \\ 
    \midrule
    Full Model (Ours) & \textbf{29.76} & \textbf{0.908} & \textbf{0.245} \\ 
    w/o weight ($w_s$)& 29.58 &\underline{0.906} &\underline{0.246} \\ 
    w/o loss ($L_{\text{reg}}$) & \underline{29.72} &\underline{0.906} & \underline{0.246} \\ 
    w/o weight \& loss& 29.57 &0.905 & 0.252 \\ 
    \bottomrule
    \end{tabular}
    }
\end{wraptable}

\begin{figure}[h!]
\centering
\includegraphics[width=\textwidth]{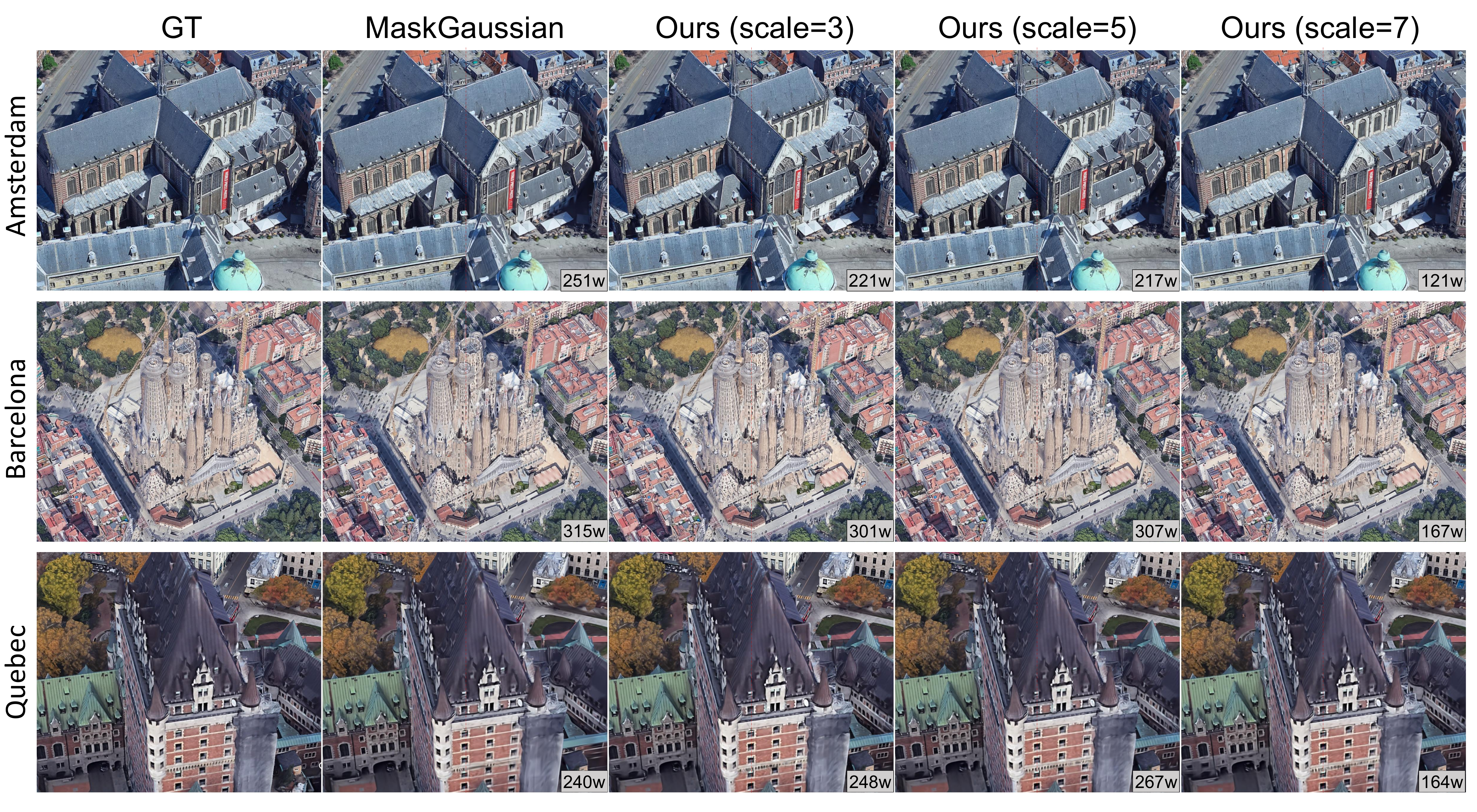}
\caption{Visual comparison of DLoD vs. CLoD strategies. The DLoD approach (the second column) uses two separate models, causing a visible quality jump at the boundary (red dashed line). Our CLoD approach (the left three columns) uses a single model with varying scale factors, resulting in a smooth, artifact-free transition.}
\label{fig:lod_comp}
\end{figure}

\begin{figure}[h!]
\centering
\includegraphics[width=\textwidth]{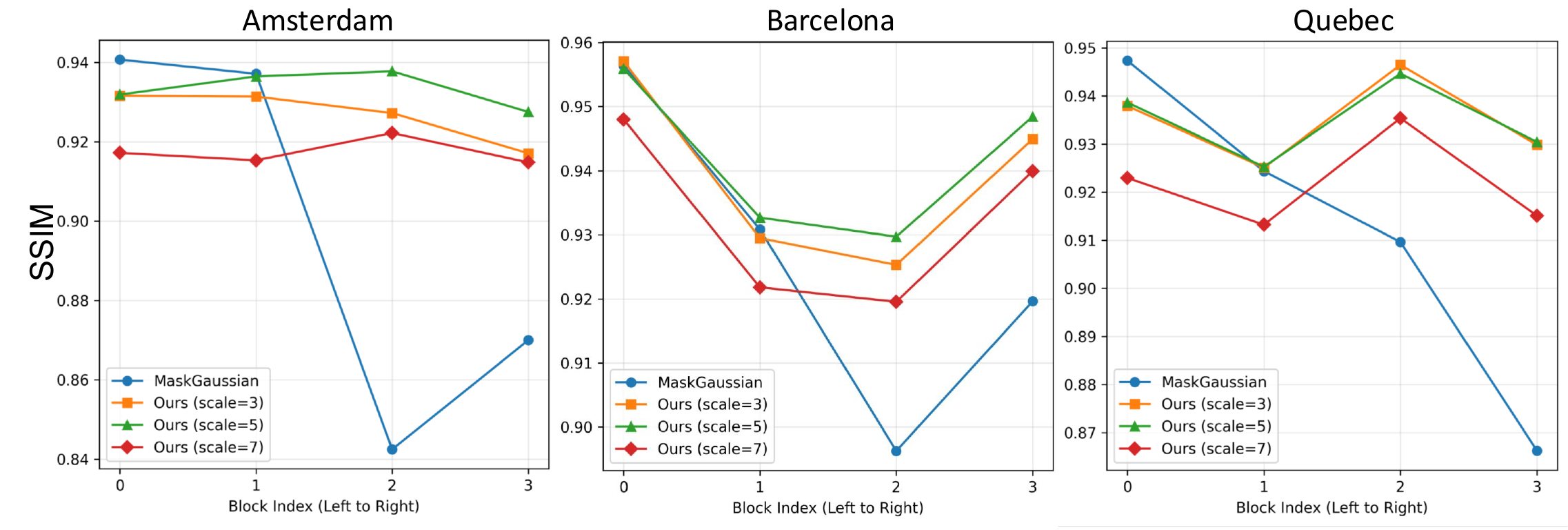}
\caption{Metric curves for the DLoD vs. CLoD comparison. The DLoD strategy exhibits a sharp, discontinuous jump in quality, whereas our CLoD strategy shows a smooth progression.}
\label{fig:lod_metric}
\end{figure}

To directly compare our continuous LoD (CLoD) approach against a traditional discrete LoD (DLoD) strategy, we designed a specific experiment. As shown in Figure~\ref{fig:lod_comp}, we divide the rendered image into four vertical regions. For the DLoD approach, we train two separate models: a high-quality baseline and a low-quality compressed model (using MaskGaussian with $\lambda_m=0.02$). We render the two left regions with the low-quality model and the two right regions with the high-quality model, simulating a hard switch between LoD levels. For our CLoD approach, we use a single trained model and render the four regions with progressively increasing detail by setting the scale factor $s_v$, respectively, while keeping the number of rendered Gaussians comparable to the DLoD setup.
The visual results clearly show the drawback of the DLoD strategy: a prominent ``popping" artifact is visible at the boundary between the two models (indicated by the red dashed line), where the quality changes abruptly. In contrast, our CLoD approach provides a smooth, continuous transition across the regions. This is further quantified in Figure~\ref{fig:lod_metric}, where the metric curves for DLoD show a sharp jump at the boundary, while our CLoD method exhibits a smooth, gradual change. Importantly, our CLoD approach is also more efficient, requiring the training of only one model, which typically takes half the time needed to train the two models required for the DLoD setup.

\subsection{Ablation Studies}

To validate our design choices, we conduct ablation studies on the key components of our training strategy, with all models trained using a maximum virtual scale of $s_v=5$. As shown in Table~\ref{tab:ablation}, removing any of these components degrades performance. We analyze the effects of the regularization loss ($L_{\text{reg}}$) and adaptive weight ($w_s$). The results confirm that all three components are crucial for achieving optimal performance. The full model consistently outperforms the ablated versions, demonstrating that the combination of our regularization loss, its adaptive weighting, and the multi-scale training approach is essential for learning a robust and efficient LoD representation.

\begin{figure}[h!]
\centering
\includegraphics[width=\textwidth]{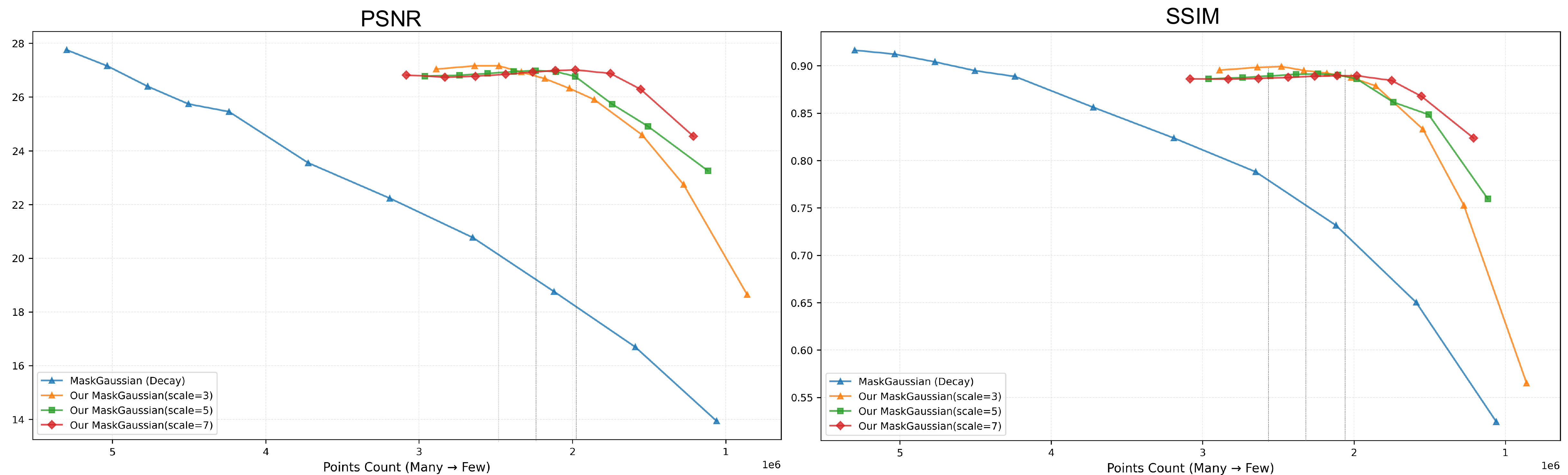}
\caption{Robustness analysis. Our CLoD-GS training strategy is applied to a MaskGaussian model on Bungeenerf dataset, successfully enabling continuous LoD on a compressed representation.}
\label{fig:robustness}
\end{figure}

\subsection{Robustness Analysis and Effectiveness Analysis}
To demonstrate the robustness and generality of our method, we apply our CLoD-GS training strategy to a model pre-trained with MaskGaussian. As shown in Figure~\ref{fig:robustness}, our method successfully imparts continuous LoD capabilities onto the already compressed MaskGaussian model. The resulting quality-compression curves show a similar trend: increasing the virtual distance scale range improves the model's ability to gracefully handle simplification. This demonstrates that our CLoD mechanism is an orthogonal enhancement that can be effectively combined with state-of-the-art compression techniques.

\begin{figure}[h!]
\centering
\includegraphics[width=0.8\textwidth]{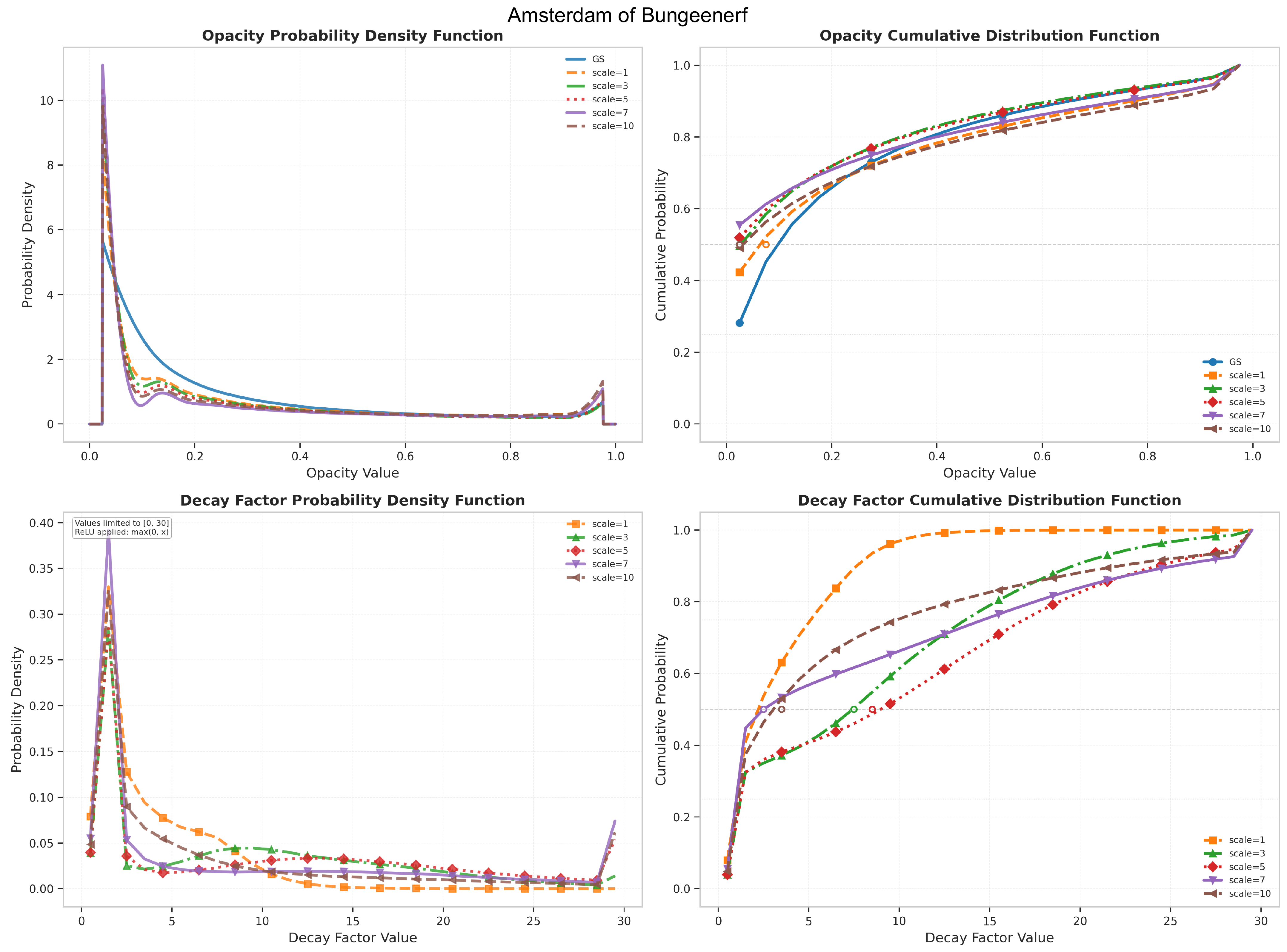}
\caption{Distribution analysis of opacity and decay factor $\sigma_{d,i}$ for various models in the Amsterdam scene. The figure plots their respective PDFs and CDFs.}
\label{fig:opa_curves}
\end{figure}

To validate the mechanism of our proposed method, we analyze the distributions of opacity and the learned decay factors ($\sigma_{d,i}$) for the Amsterdam scene from the BungeeNeRF dataset, as depicted in Figure~\ref{fig:opa_curves}. The analysis empirically demonstrates that the virtual distance range during training is the governing factor for $\sigma_{d,i}$. We observe that a wider distance range encourages a distribution of $\sigma_{d,i}$ skewed towards larger values. Primitives with larger $\sigma_{d,i}$ values exhibit slower opacity decay, thus remaining visible at greater virtual distances. This confirms that our model automatically learns to assign higher persistence to perceptually important primitives without requiring any manual supervision.
Our method adds only one additional float parameter per Gaussian. In a standard 3DGS implementation, each Gaussian requires approximately 248 bytes of storage, an increase of only 1.6\%, which is an entirely acceptable overhead.

\section{Conclusion}

In this paper, we introduced CLoD-GS, a novel framework that seamlessly integrates a CLoD mechanism into the 3DGS representation. We identified the core limitations of applying traditional DLoD paradigms to 3DGS—namely, prohibitive storage overhead and jarring visual ``popping" artifacts. Our core contribution is a lightweight, learnable mechanism that augments each Gaussian primitive with a distance-dependent decay parameter. This parameter, optimized directly within a novel virtual distance scaling training strategy, allows each primitive to learn its own view-dependent simplification behavior. The resulting model contains a continuous spectrum of detail within a single, unified representation, enabling smooth, pop-free quality scaling. Our experiments have shown that CLoD-GS not only eliminates the fundamental drawbacks of DLoD but also achieves high rendering quality, often with a more compact set of primitives than the baseline 3DGS.
While our method demonstrates significant advantages for static scenes, future work could explore the integration of more sophisticated perceptual metrics beyond distance to guide the learned decay, or hybrid systems that combine our continuous per-primitive scaling with chunk-based loading for rendering massive-scale environments. Ultimately, CLoD-GS represents a significant step towards creating truly scalable, efficient, and visually coherent neural scene representations, paving the way for the next generation of real-time, high-fidelity graphics applications.

\section{Acknowledgement}
This study was supported by National Key Research and Development Program of China (2022YFC3803002), and the Xplorer Prize.

\section{Reproducibility Statement}
We are committed to ensuring the reproducibility of our research. To this end, we provide comprehensive resources to facilitate the verification of our findings. All experiments were conducted on publicly available datasets, ensuring that the data is accessible to the broader research community. The detailed experimental setup, including all hyperparameters and training configurations, is described in Section~\ref{sec:experimets} of the main paper.

\bibliography{main}
\bibliographystyle{iclr2026/iclr2026_conference}

\appendix
\section{Rendering Result on MipNeRF360 Dataset}

We have added experiments on the challenging Mip-NeRF 360 dataset. As shown in the Table ~\ref{tab:360_comparison}, even without any dataset-specific tuning, CLoD-GS demonstrates a competitive trade-off between quality and memory. We acknowledge that the results are not state-of-the-art on this specific dataset, which we attribute to our use of universal parameters that were not optimized for these special scenes. However, this finding underscores a key strength of our method: generality. In contrast, other methods like Octree-GS often require dataset-specific configurations, as evidenced by the separate training scripts provided in its official repository for different datasets.

\begin{table}[htbp]
  \centering
  \caption{Quantitative comparison of different methods on the \textbf{MipNeRF360 dataset}. The arrows indicate whether higher (↑) or lower (↓) values are better.}
  \label{tab:360_comparison}
  \begin{tabular}{lrrrrr}
    \toprule
    Method & PSNR↑ & SSIM↑ & LPIPS↓ & \#GS (k)↓ & Memory (MB)↓ \\
    \midrule
    3DGS           & 27.59 & 0.8136 & 0.2206 & 2638.53 & 624.04 \\
    Fast Rendering & 27.58 & 0.8240 & 0.2420 & 2638.53 & 624.04 \\
    Octree-GS      & 27.65 & 0.8150 & 0.2200 & /       & 418.60 \\
    MaskGaussian   & 27.23 & 0.8035 & 0.2253 & 2145.10 & 507.34 \\
    \midrule
    Ours (scale=1) & 27.01 & 0.8069 & 0.2299 & 2414.38 & 580.24 \\
    Ours (scale=3) & 26.99 & 0.8042 & 0.2361 & 1778.10 & 427.32 \\
    Ours (scale=7) & 26.71 & 0.7968 & 0.2494 & 1361.49 & 327.20 \\
    \bottomrule
  \end{tabular}
\end{table}

\section{Detailed Metric Tables}
This section provides a comprehensive, scene-by-scene breakdown of the performance metrics for the methods evaluated in our study. These tables offer a granular view of the results that are summarized in the main body of the paper, detailing the PSNR, SSIM, LPIPS, and Gaussian count for our method, the original 3DGS, MaskGaussian, and the ablation study configurations across all tested datasets.
\label{app:appendix}
\begin{table}[htbp]
  \centering
  \caption{ours (scale=1)}
  \label{tab:our_method_results}
  \begin{tabular}{lcccc}
    \toprule
    Dataset & PSNR↑ & SSIM↑ & LPIPS↓ & \#GS (M)↓ \\
    \midrule
    \multicolumn{5}{c}{\textit{BungeeNeRF}} \\
    \midrule
    amsterdam & 27.78 & 0.911 & 0.108 & 3.89 \\
    barcelona & 27.84 & 0.922 & 0.085 & 4.91 \\
    bilbao    & 29.02 & 0.919 & 0.101 & 3.63 \\
    chicago   & 28.62 & 0.934 & 0.087 & 3.50 \\
    hollywood & 26.53 & 0.880 & 0.138 & 4.58 \\
    pompidou  & 27.37 & 0.920 & 0.094 & 5.14 \\
    quebec    & 29.03 & 0.938 & 0.092 & 3.75 \\
    rome      & 28.23 & 0.925 & 0.098 & 4.08 \\
    \midrule
    \multicolumn{5}{c}{\textit{Deep Blending}} \\
    \midrule
    drjohnson & 29.47 & 0.905 & 0.237 & 2.07 \\
    playroom  & 30.39 & 0.911 & 0.241 & 1.33 \\
    \midrule
    \multicolumn{5}{c}{\textit{Tanks \& Temples}} \\
    \midrule
    train     & 22.14 & 0.806 & 0.220 & 0.57 \\
    truck     & 25.35 & 0.880 & 0.150 & 1.75 \\
    \bottomrule
  \end{tabular}
\end{table}

\begin{table}[htbp]
  \centering
  \caption{Mask Gaussian }
  \label{tab:mask_gaussian_results}
  \begin{tabular}{lcccc}
    \toprule
    Dataset & PSNR↑ & SSIM↑ & LPIPS↓ & \#GS (M)↓ \\
    \midrule
    \multicolumn{5}{c}{\textit{BungeeNeRF}} \\
    \midrule
    amsterdam & 27.69 & 0.916 & 0.100 & 4.69 \\
    barcelona & 27.49 & 0.919 & 0.085 & 6.62 \\
    bilbao    & 28.86 & 0.919 & 0.097 & 4.28 \\
    chicago   & 28.03 & 0.931 & 0.085 & 4.75 \\
    hollywood & 26.35 & 0.873 & 0.133 & 5.62 \\
    pompidou  & 27.13 & 0.920 & 0.093 & 6.83 \\
    quebec    & 28.84 & 0.936 & 0.092 & 4.44 \\
    rome      & 27.66 & 0.918 & 0.100 & 5.17 \\
    \midrule
    \multicolumn{5}{c}{\textit{Deep Blending}} \\
    \midrule
    drjohnson & 29.21 & 0.904 & 0.243 & 2.34 \\
    playroom  & 30.12 & 0.910 & 0.244 & 1.22 \\
    \midrule
    \multicolumn{5}{c}{\textit{Tanks \& Temples}} \\
    \midrule
    train     & 21.80 & 0.811 & 0.212 & 0.88 \\
    truck     & 25.31 & 0.881 & 0.149 & 1.60 \\
    \bottomrule
  \end{tabular}
\end{table}

\begin{table}[htbp]
  \centering
  \caption{Original 3DGS}
  \label{tab:original_gs_results}
  \begin{tabular}{lcccc}
    \toprule
    Dataset & PSNR↑ & SSIM↑ & LPIPS↓ & \#GS (M)↓ \\
    \midrule
    \multicolumn{5}{c}{\textit{BungeeNeRF}} \\
    \midrule
    amsterdam & 27.85 & 0.918 & 0.096 & 6.24 \\
    barcelona & 27.67 & 0.920 & 0.083 & 8.18 \\
    bilbao    & 28.98 & 0.918 & 0.095 & 5.49 \\
    chicago   & 28.54 & 0.933 & 0.080 & 6.08 \\
    hollywood & 26.24 & 0.868 & 0.135 & 6.79 \\
    pompidou  & 27.26 & 0.921 & 0.091 & 8.64 \\
    quebec    & 28.95 & 0.937 & 0.089 & 5.91 \\
    rome      & 27.77 & 0.920 & 0.096 & 6.55 \\
    \midrule
    \multicolumn{5}{c}{\textit{Deep Blending}} \\
    \midrule
    drjohnson & 29.43 & 0.905 & 0.236 & 3.12 \\
    playroom  & 30.25 & 0.909 & 0.240 & 1.85 \\
    \midrule
    \multicolumn{5}{c}{\textit{Tanks \& Temples}} \\
    \midrule
    train     & 21.97 & 0.821 & 0.197 & 1.09 \\
    truck     & 25.44 & 0.885 & 0.142 & 2.06 \\
    \bottomrule
  \end{tabular}
\end{table}

\begin{table}[htbp]
  \centering
  \caption{Full Model (scale=5)}
  \label{tab:ablation_full}
  \begin{tabular}{lcccc}
    \toprule
    Dataset & PSNR↑ & SSIM↑ & LPIPS↓ & \#GS (M)↓ \\
    \midrule
    \multicolumn{5}{c}{\textit{BungeeNeRF}} \\
    \midrule
    amsterdam & 27.15 & 0.889 & 0.137 & 2.37 \\
    barcelona & 26.99 & 0.886 & 0.123 & 2.78 \\
    bilbao    & 28.62 & 0.905 & 0.122 & 2.22 \\
    chicago   & 28.01 & 0.917 & 0.108 & 2.30 \\
    hollywood & 26.75 & 0.878 & 0.147 & 3.34 \\
    pompidou  & 26.85 & 0.902 & 0.118 & 2.92 \\
    quebec    & 28.89 & 0.932 & 0.105 & 2.53 \\
    rome      & 27.43 & 0.903 & 0.127 & 2.16 \\
    \midrule
    \multicolumn{5}{c}{\textit{Deep Blending}} \\
    \midrule
    drjohnson & 29.32 & 0.906 & 0.241 & 1.16 \\
    playroom  & 30.20 & 0.909 & 0.248 & 0.86 \\
    \midrule
    \multicolumn{5}{c}{\textit{Tanks \& Temples}} \\
    \midrule
    train     & 21.99 & 0.806 & 0.219 & 0.59 \\
    truck     & 25.29 & 0.880 & 0.148 & 1.46 \\
    \bottomrule
  \end{tabular}
\end{table}

\begin{table}[htbp]
  \centering
  \caption{Without Weight Adaptation (scale=5)}
  \label{tab:ablation_noweight}
  \begin{tabular}{lcccc}
    \toprule
    Dataset & PSNR↑ & SSIM↑ & LPIPS↓ & \#GS (M)↓ \\
    \midrule
    \multicolumn{5}{c}{\textit{BungeeNeRF}} \\
    \midrule
    amsterdam & 26.82 & 0.876 & 0.146 & 2.34 \\
    barcelona & 26.33 & 0.863 & 0.138 & 3.11 \\
    bilbao    & 28.27 & 0.896 & 0.128 & 2.07 \\
    chicago   & 27.82 & 0.913 & 0.111 & 2.17 \\
    hollywood & 26.75 & 0.882 & 0.140 & 3.65 \\
    pompidou  & 26.86 & 0.895 & 0.121 & 3.10 \\
    quebec    & 28.90 & 0.933 & 0.103 & 2.40 \\
    rome      & 27.32 & 0.895 & 0.130 & 2.29 \\
    \midrule
    \multicolumn{5}{c}{\textit{Deep Blending}} \\
    \midrule
    drjohnson & 29.15 & 0.905 & 0.242 & 1.12 \\
    playroom  & 30.01 & 0.907 & 0.250 & 0.92 \\
    \midrule
    \multicolumn{5}{c}{\textit{Tanks \& Temples}} \\
    \midrule
    train     & 22.02 & 0.809 & 0.216 & 0.63 \\
    truck     & 25.26 & 0.881 & 0.147 & 1.46 \\
    \bottomrule
  \end{tabular}
\end{table}

\begin{table}[htbp]
  \centering
  \caption{Without Regularization (scale=5)}
  \label{tab:ablation_noreg}
  \begin{tabular}{lcccc}
    \toprule
    Dataset & PSNR↑ & SSIM↑ & LPIPS↓ & \#GS (M)↓ \\
    \midrule
    \multicolumn{5}{c}{\textit{BungeeNeRF}} \\
    \midrule
    amsterdam & 27.09 & 0.888 & 0.138 & 2.37 \\
    barcelona & 27.10 & 0.893 & 0.119 & 2.85 \\
    bilbao    & 28.51 & 0.906 & 0.121 & 2.29 \\
    chicago   & 27.98 & 0.917 & 0.108 & 2.29 \\
    hollywood & 26.67 & 0.878 & 0.147 & 3.42 \\
    pompidou  & 26.95 & 0.906 & 0.116 & 2.91 \\
    quebec    & 28.86 & 0.932 & 0.105 & 2.56 \\
    rome      & 27.34 & 0.899 & 0.129 & 2.14 \\
    \midrule
    \multicolumn{5}{c}{\textit{Deep Blending}} \\
    \midrule
    drjohnson & 29.28 & 0.906 & 0.242 & 1.12 \\
    playroom  & 30.17 & 0.907 & 0.250 & 0.93 \\
    \midrule
    \multicolumn{5}{c}{\textit{Tanks \& Temples}} \\
    \midrule
    train     & 22.10 & 0.808 & 0.218 & 0.60 \\
    truck     & 25.22 & 0.880 & 0.149 & 1.38 \\
    \bottomrule
  \end{tabular}
\end{table}

\begin{table}[htbp]
  \centering
  \caption{Without Weight \& Regularization (scale=5)}
  \label{tab:ablation_noweight_noreg}
  \begin{tabular}{lcccc}
    \toprule
    Dataset & PSNR↑ & SSIM↑ & LPIPS↓ & \#GS (M)↓ \\
    \midrule
    \multicolumn{5}{c}{\textit{BungeeNeRF}} \\
    \midrule
    amsterdam & 25.93 & 0.848 & 0.193 & 1.03 \\
    barcelona & 26.04 & 0.849 & 0.171 & 1.37 \\
    bilbao    & 27.97 & 0.884 & 0.158 & 1.06 \\
    chicago   & 27.18 & 0.893 & 0.147 & 0.81 \\
    hollywood & 26.02 & 0.843 & 0.208 & 1.46 \\
    pompidou  & 26.08 & 0.877 & 0.153 & 1.37 \\
    quebec    & 28.05 & 0.912 & 0.140 & 1.06 \\
    rome      & 26.46 & 0.866 & 0.179 & 0.97 \\
    \midrule
    \multicolumn{5}{c}{\textit{Deep Blending}} \\
    \midrule
    drjohnson & 29.31 & 0.905 & 0.248 & 0.94 \\
    playroom  & 29.84 & 0.904 & 0.257 & 0.63 \\
    \midrule
    \multicolumn{5}{c}{\textit{Tanks \& Temples}} \\
    \midrule
    train     & 22.30 & 0.800 & 0.236 & 0.38 \\
    truck     & 25.21 & 0.876 & 0.158 & 1.09 \\
    \bottomrule
  \end{tabular}
\end{table}

\section{Detailed Metric Curves}

This section presents the detailed quality-versus-performance curves for each of the eight scenes in the BungeeNeRF dataset. The following figures illustrate the trade-off between PSNR and the number of rendered primitives for both the original 3DGS and the MaskGaussian methods, providing a visual complement to the quantitative data in the preceding tables.

\begin{figure}[h!]
\centering
\includegraphics[width=\textwidth]{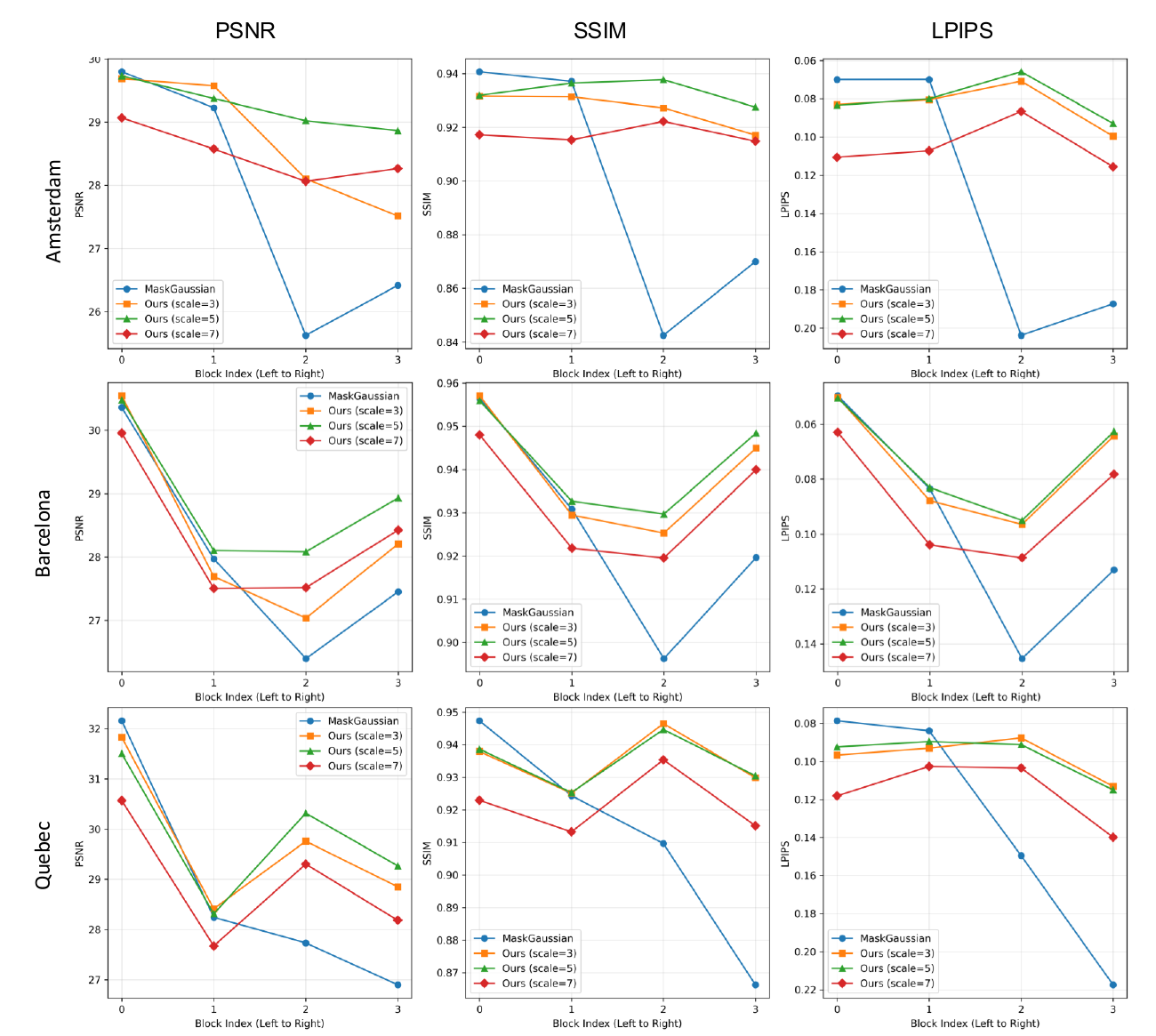}
\caption{All metric curves for the DLoD vs. CLoD comparison.}
\label{fig:lod_metric_all}
\end{figure}

\begin{figure}[h!]
\centering
\includegraphics[width=\textwidth]{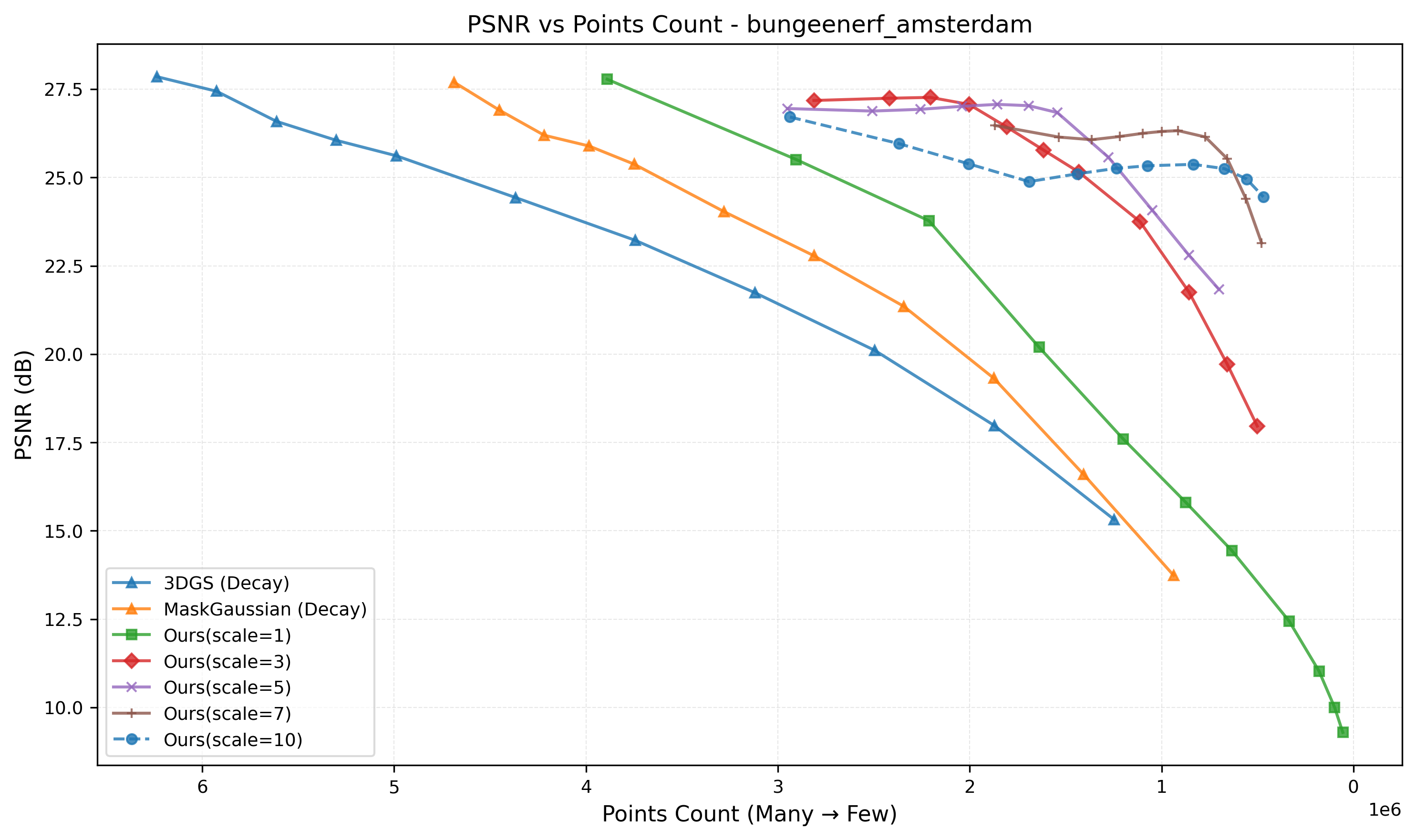}
\caption{3DGS: PSNR vs. primitive count on the BungeeNeRF amsterdam dataset.}
\label{fig:curve_3dgs_amsterdam}
\end{figure}

\begin{figure}[h!]
\centering
\includegraphics[width=\textwidth]{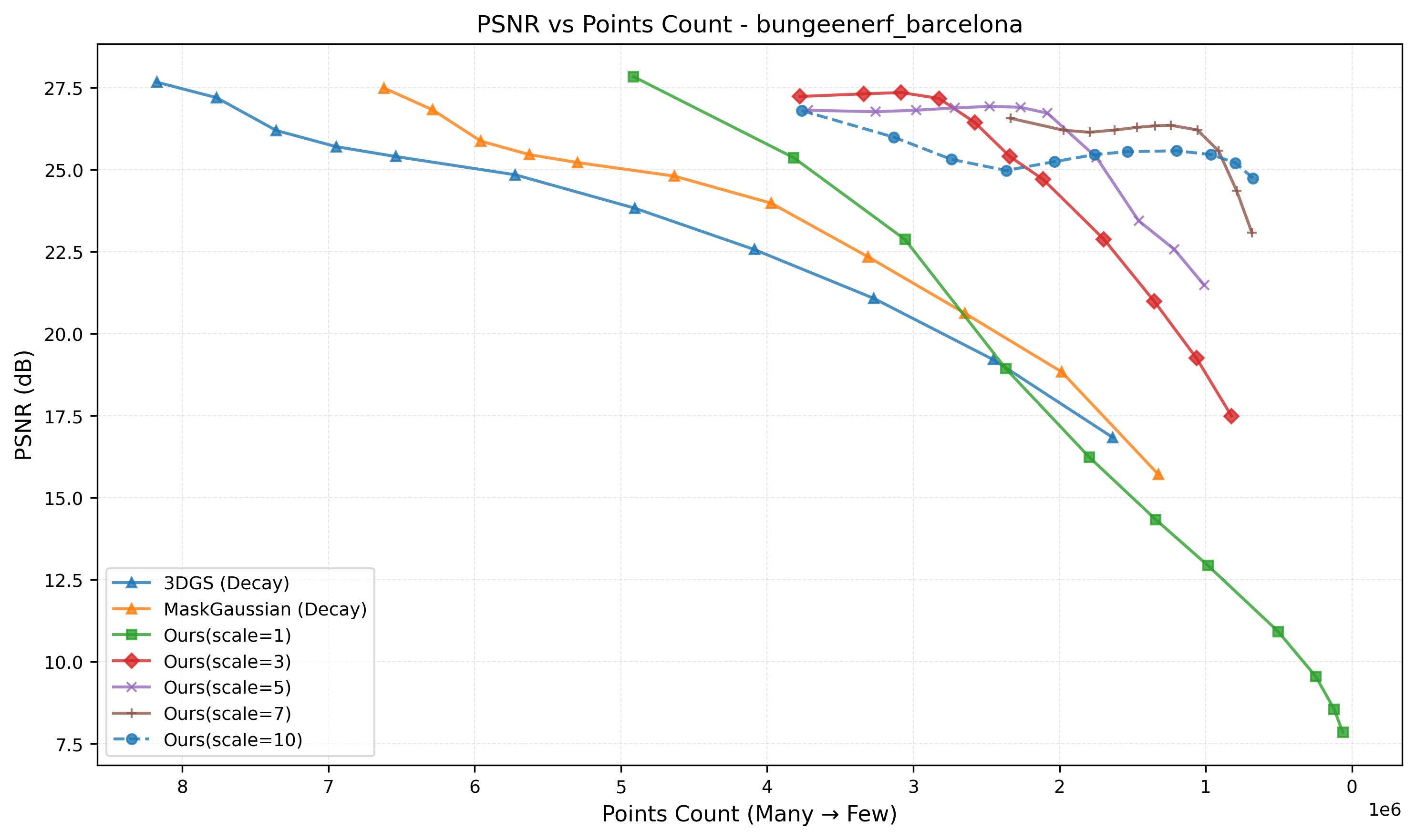}
\caption{3DGS: PSNR vs. primitive count on the BungeeNeRF barcelona dataset.}
\label{fig:curve_3dgs_barcelona}
\end{figure}

\begin{figure}[h!]
\centering
\includegraphics[width=\textwidth]{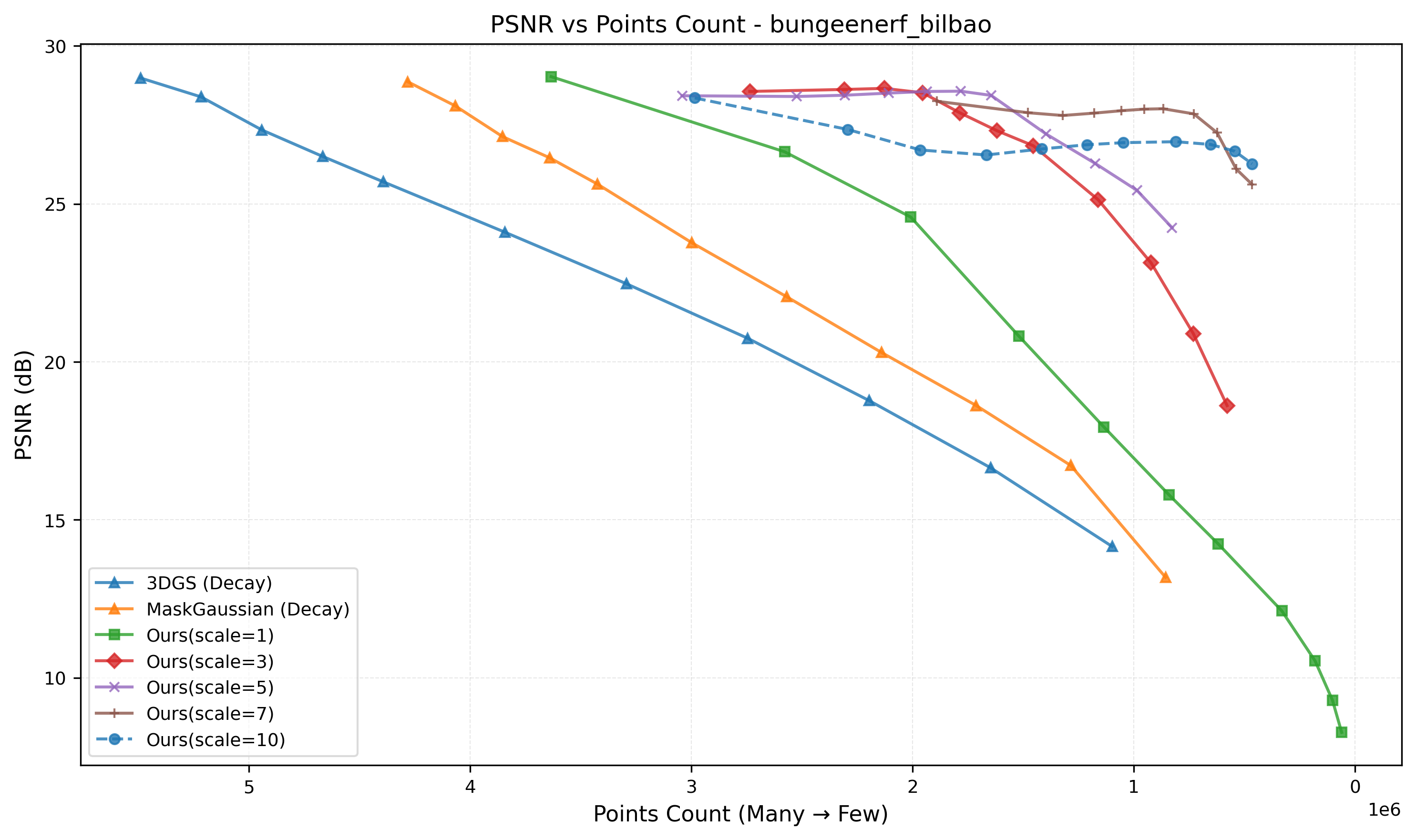}
\caption{3DGS: PSNR vs. primitive count on the BungeeNeRF bilbao dataset.}
\label{fig:curve_3dgs_bilbao}
\end{figure}

\begin{figure}[h!]
\centering
\includegraphics[width=\textwidth]{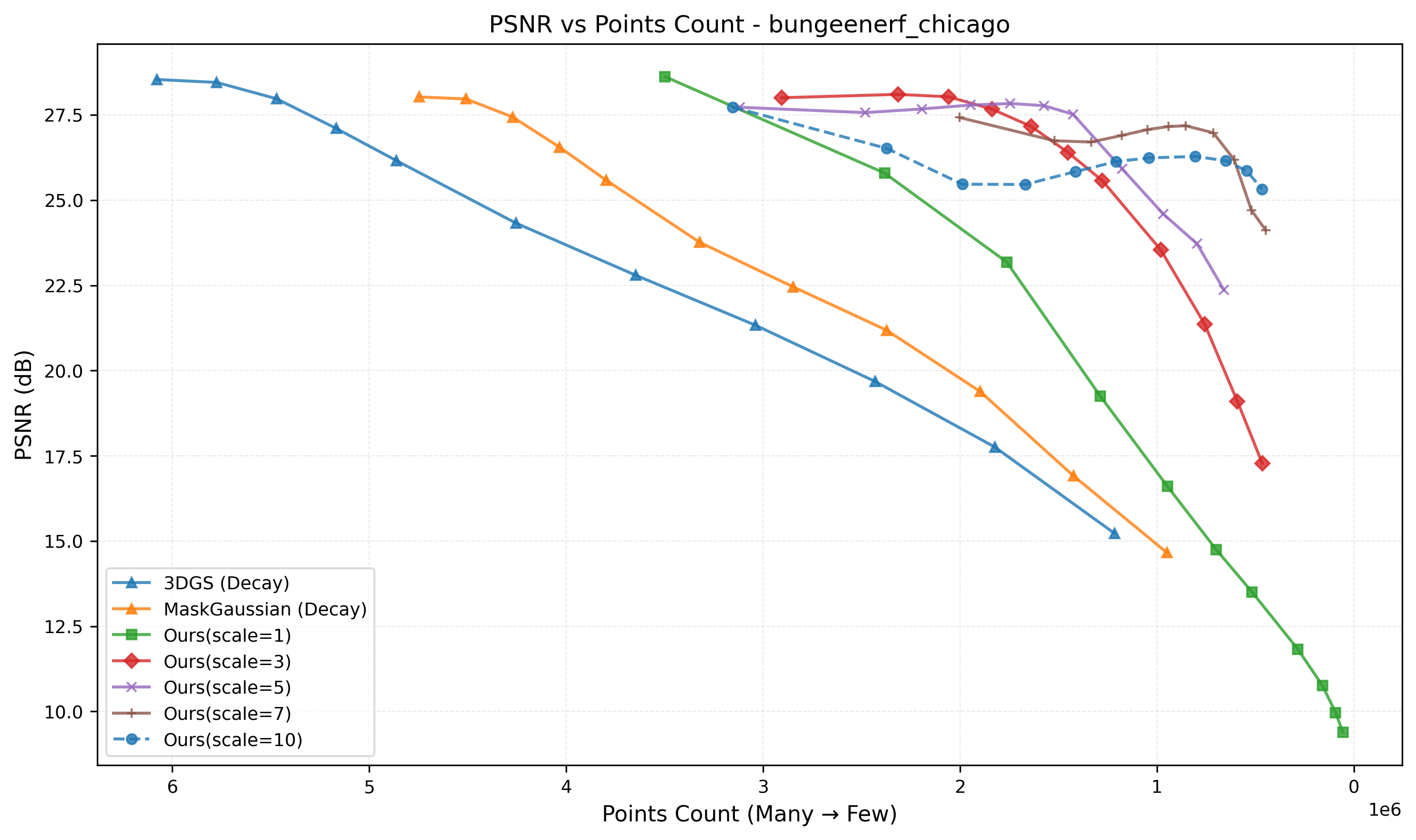}
\caption{3DGS: PSNR vs. primitive count on the BungeeNeRF chicago dataset.}
\label{fig:curve_3dgs_chicago}
\end{figure}

\begin{figure}[h!]
\centering
\includegraphics[width=\textwidth]{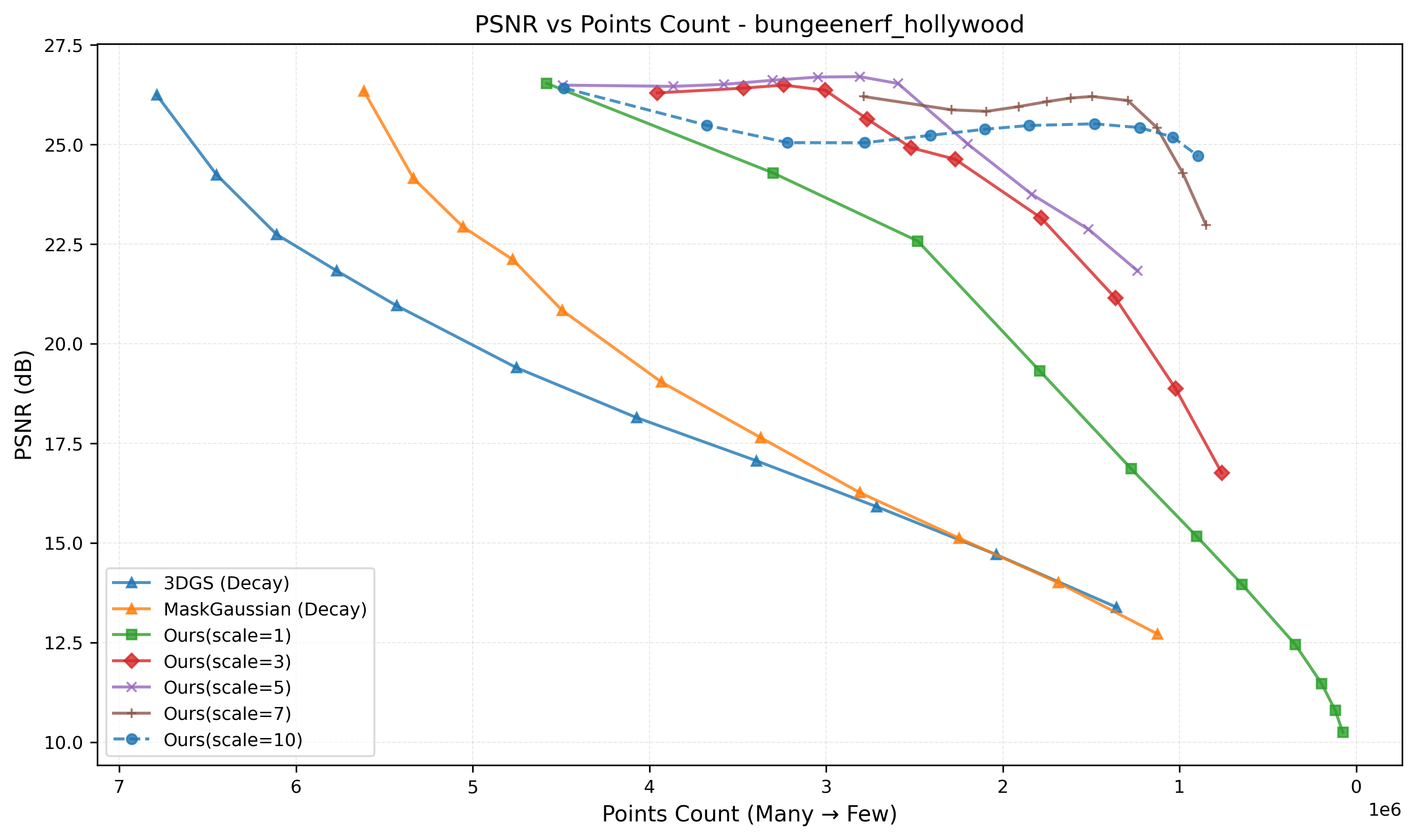}
\caption{3DGS: PSNR vs. primitive count on the BungeeNeRF hollywood dataset.}
\label{fig:curve_3dgs_hollywood}
\end{figure}

\begin{figure}[h!]
\centering
\includegraphics[width=\textwidth]{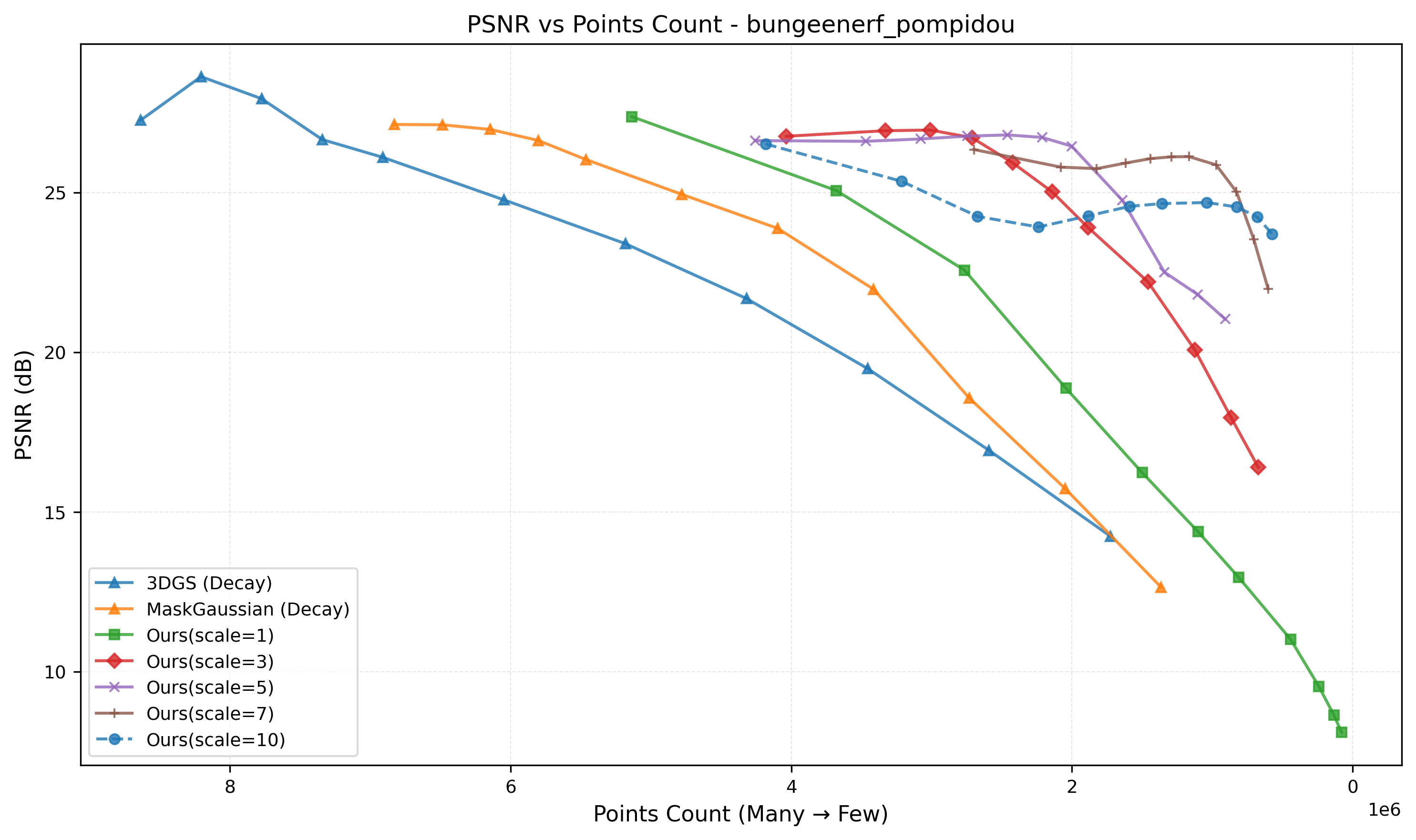}
\caption{3DGS: PSNR vs. primitive count on the BungeeNeRF pompidou dataset.}
\label{fig:curve_3dgs_pompidou}
\end{figure}

\begin{figure}[h!]
\centering
\includegraphics[width=\textwidth]{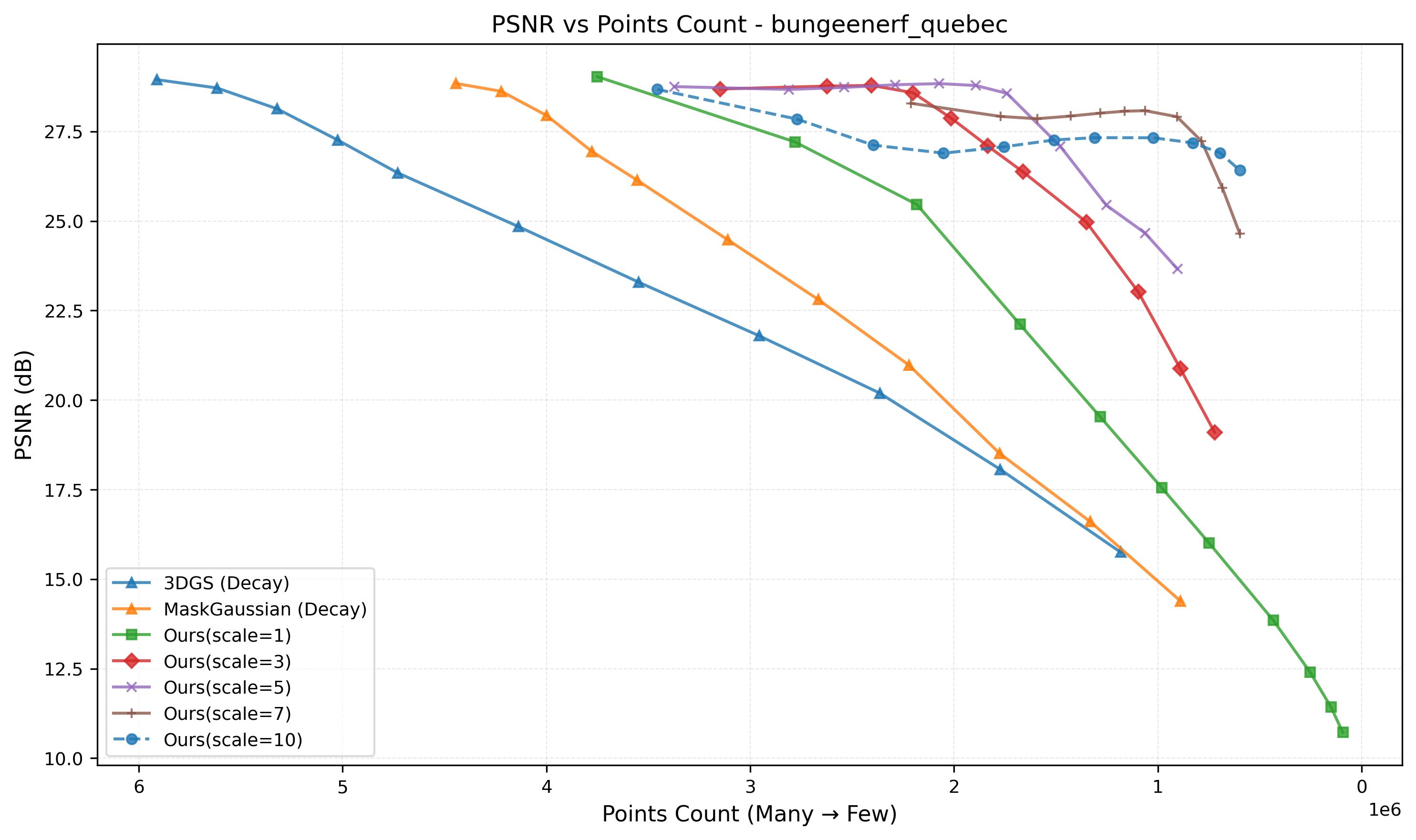}
\caption{3DGS: PSNR vs. primitive count on the BungeeNeRF quebec dataset.}
\label{fig:curve_3dgs_quebec}
\end{figure}

\begin{figure}[h!]
\centering
\includegraphics[width=\textwidth]{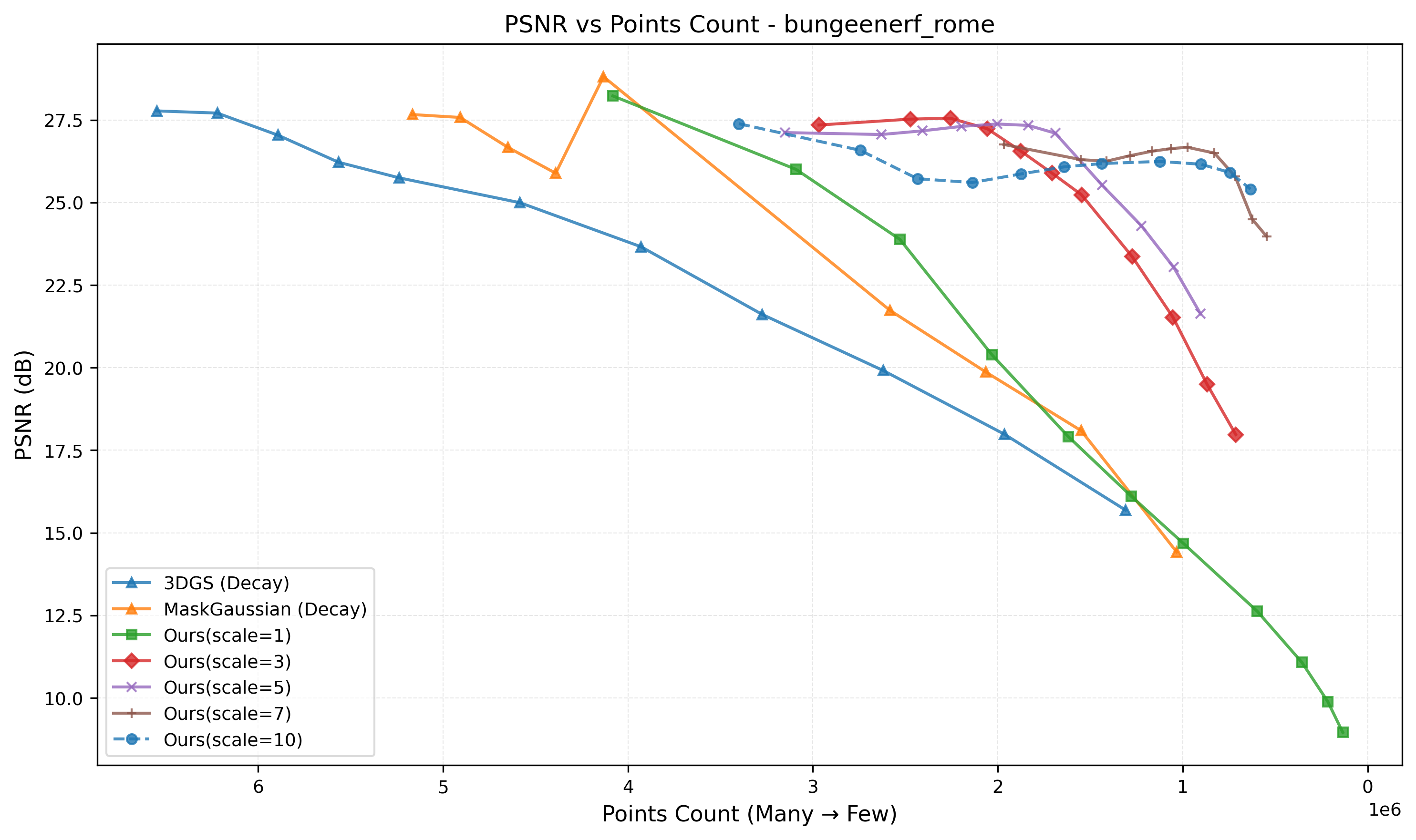}
\caption{3DGS: PSNR vs. primitive count on the BungeeNeRF rome dataset.}
\label{fig:curve_3dgs_rome}
\end{figure}

\begin{figure}[h!]
\centering
\includegraphics[width=\textwidth]{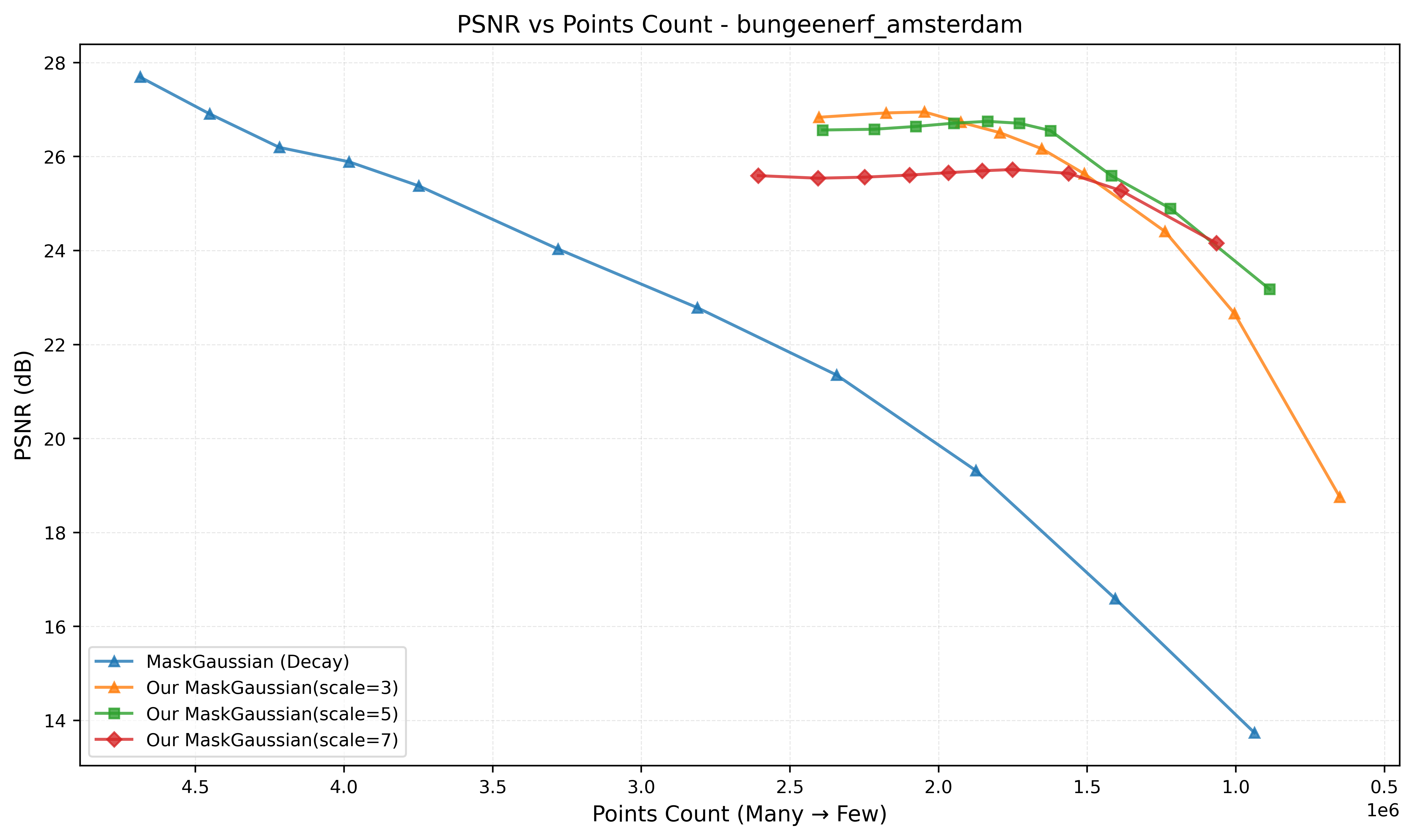}
\caption{MaskGaussian: PSNR vs. primitive count on the BungeeNeRF amsterdam dataset.}
\label{fig:curve_mask_amsterdam}
\end{figure}

\begin{figure}[h!]
\centering
\includegraphics[width=\textwidth]{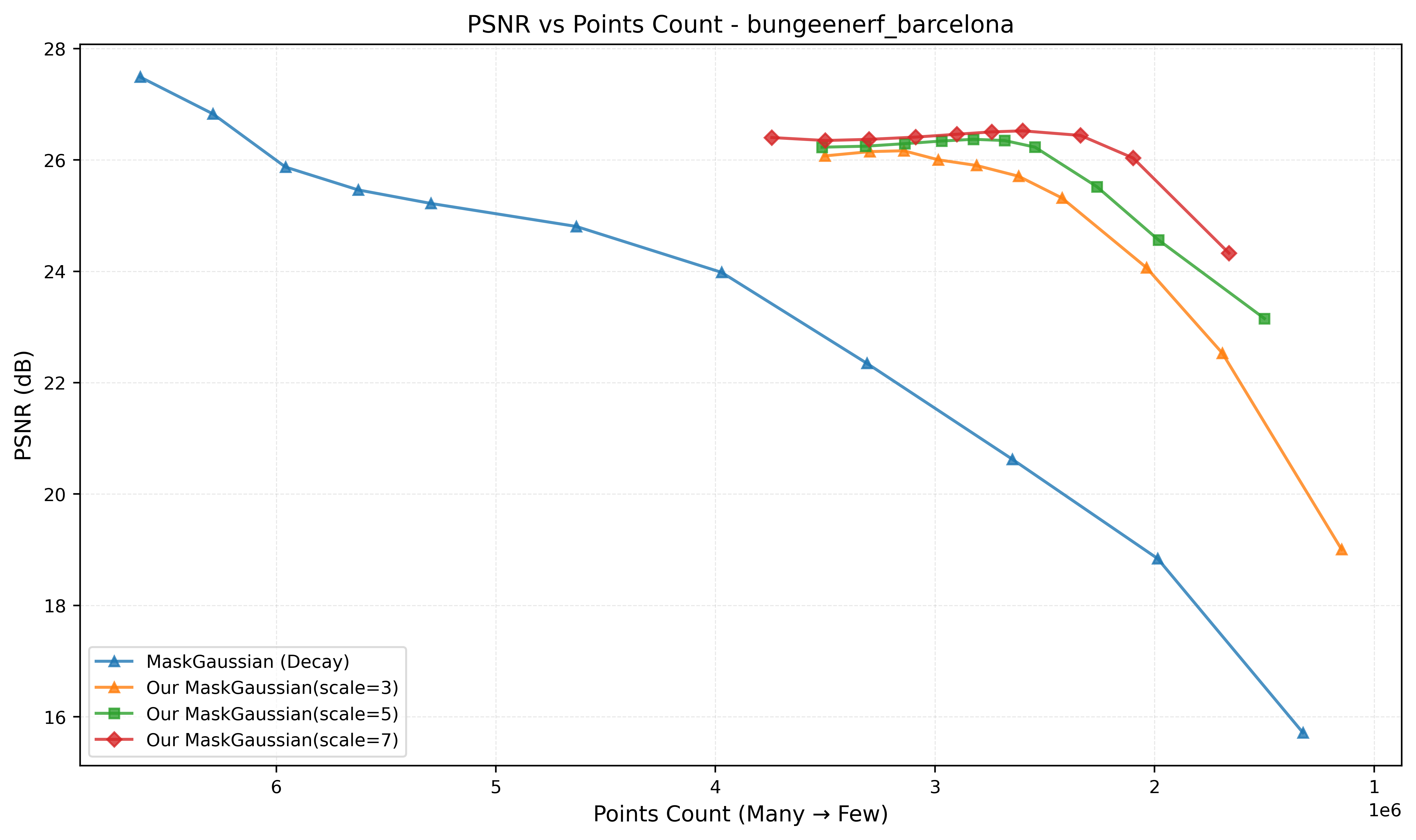}
\caption{MaskGaussian: PSNR vs. primitive count on the BungeeNeRF barcelona dataset.}
\label{fig:curve_mask_barcelona}
\end{figure}

\begin{figure}[h!]
\centering
\includegraphics[width=\textwidth]{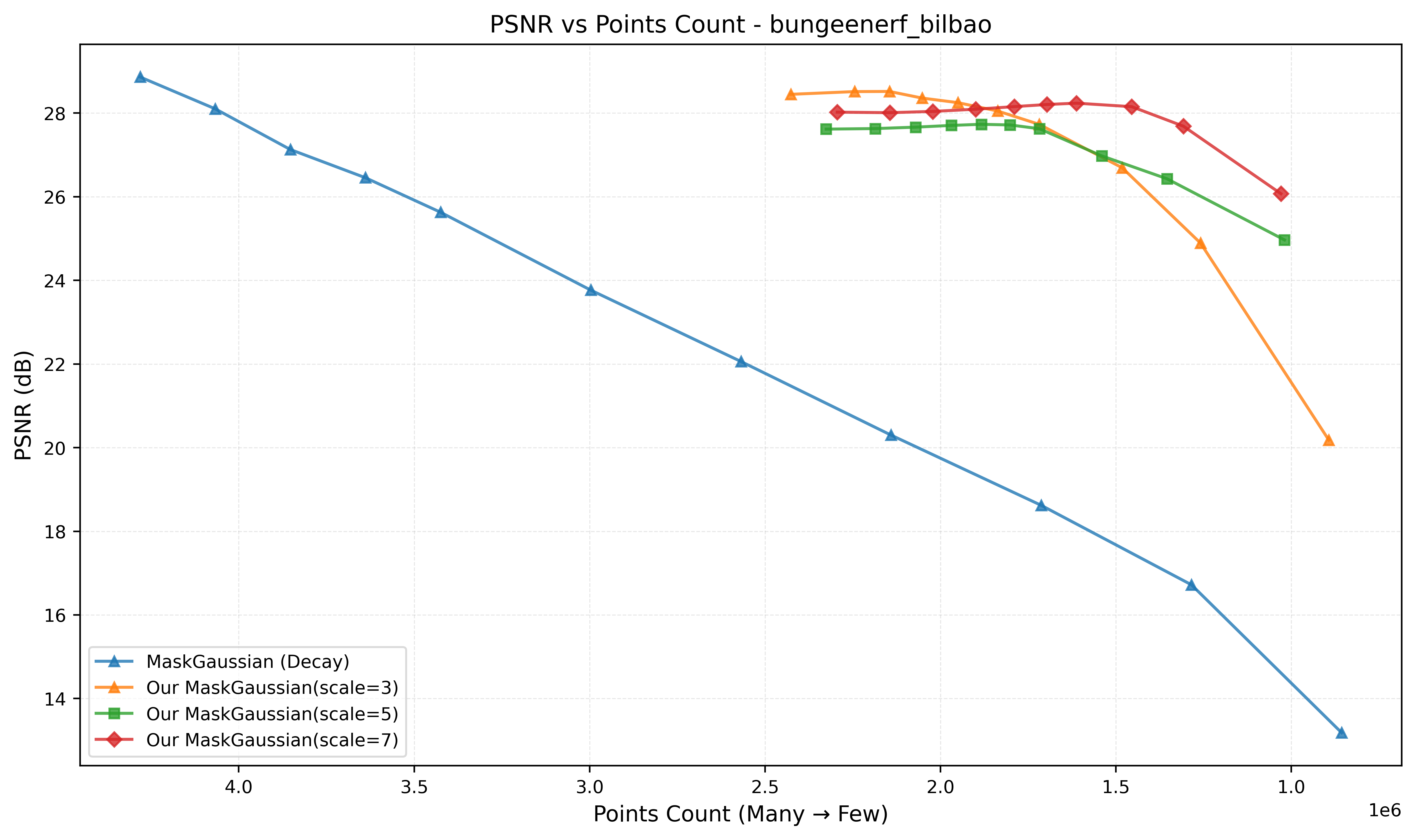}
\caption{MaskGaussian: PSNR vs. primitive count on the BungeeNeRF bilbao dataset.}
\label{fig:curve_mask_bilbao}
\end{figure}

\begin{figure}[h!]
\centering
\includegraphics[width=\textwidth]{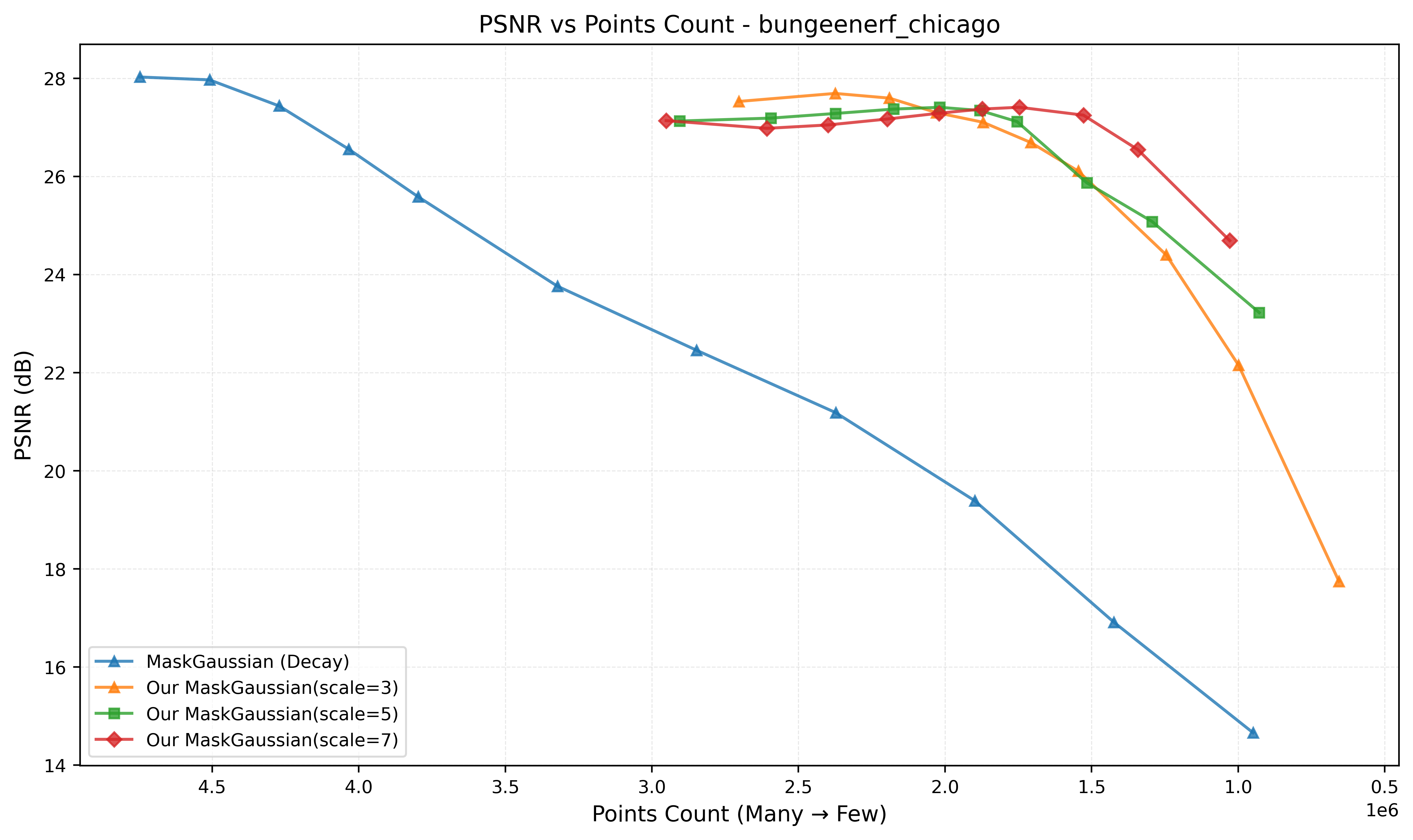}
\caption{MaskGaussian: PSNR vs. primitive count on the BungeeNeRF chicago dataset.}
\label{fig:curve_mask_chicago}
\end{figure}

\begin{figure}[h!]
\centering
\includegraphics[width=\textwidth]{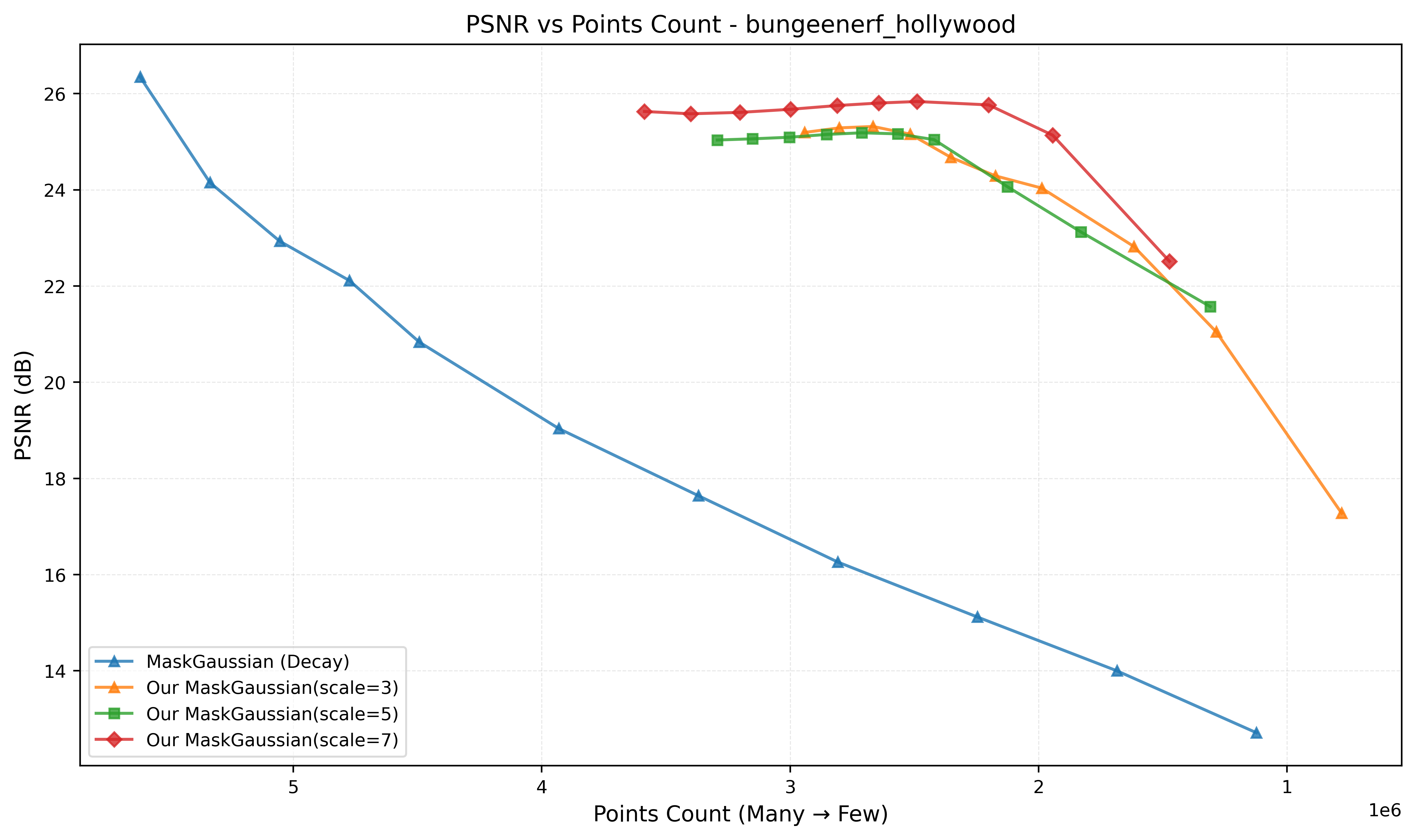}
\caption{MaskGaussian: PSNR vs. primitive count on the BungeeNeRF hollywood dataset.}
\label{fig:curve_mask_hollywood}
\end{figure}

\begin{figure}[h!]
\centering
\includegraphics[width=\textwidth]{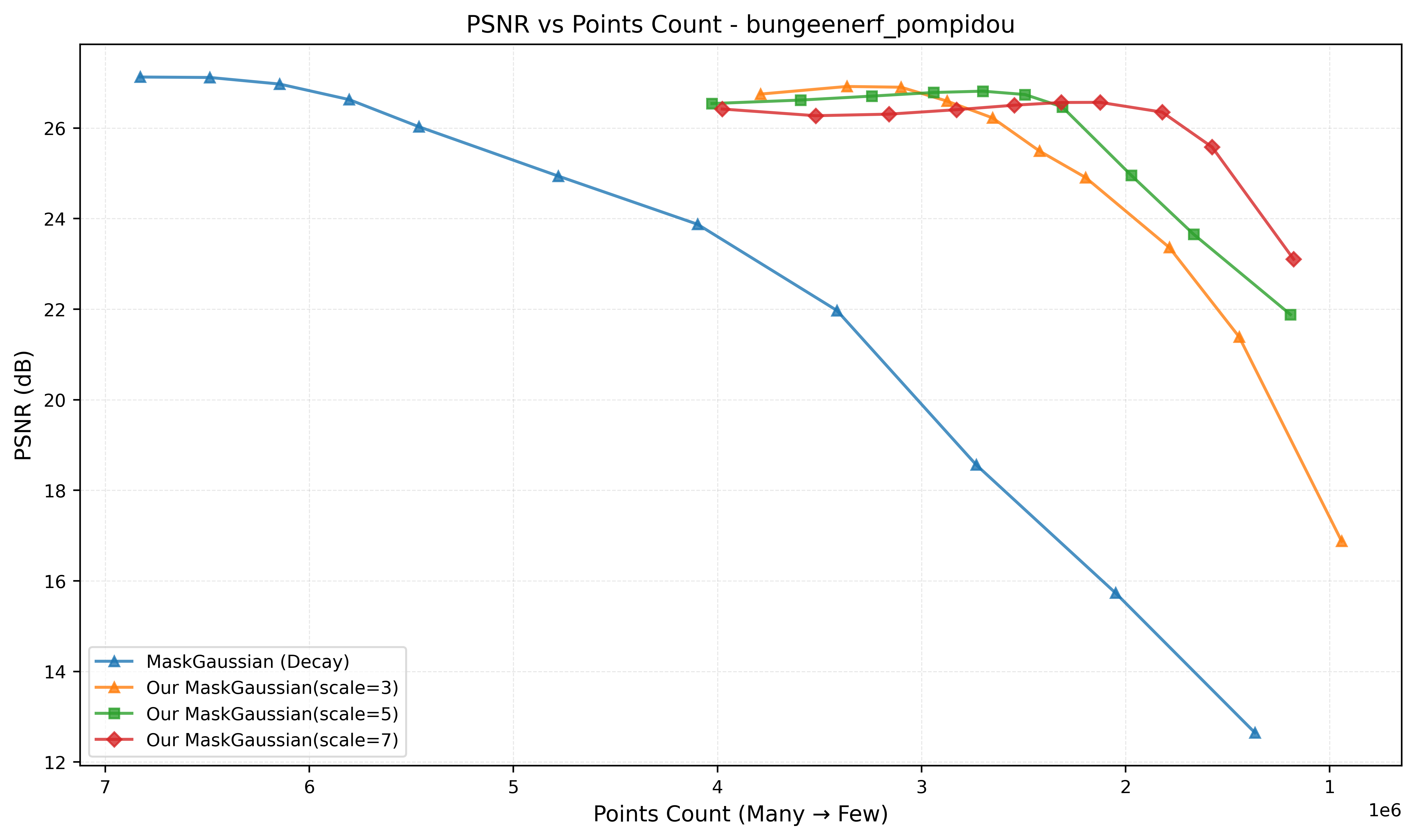}
\caption{MaskGaussian: PSNR vs. primitive count on the BungeeNeRF pompidou dataset.}
\label{fig:curve_mask_pompidou}
\end{figure}

\begin{figure}[h!]
\centering
\includegraphics[width=\textwidth]{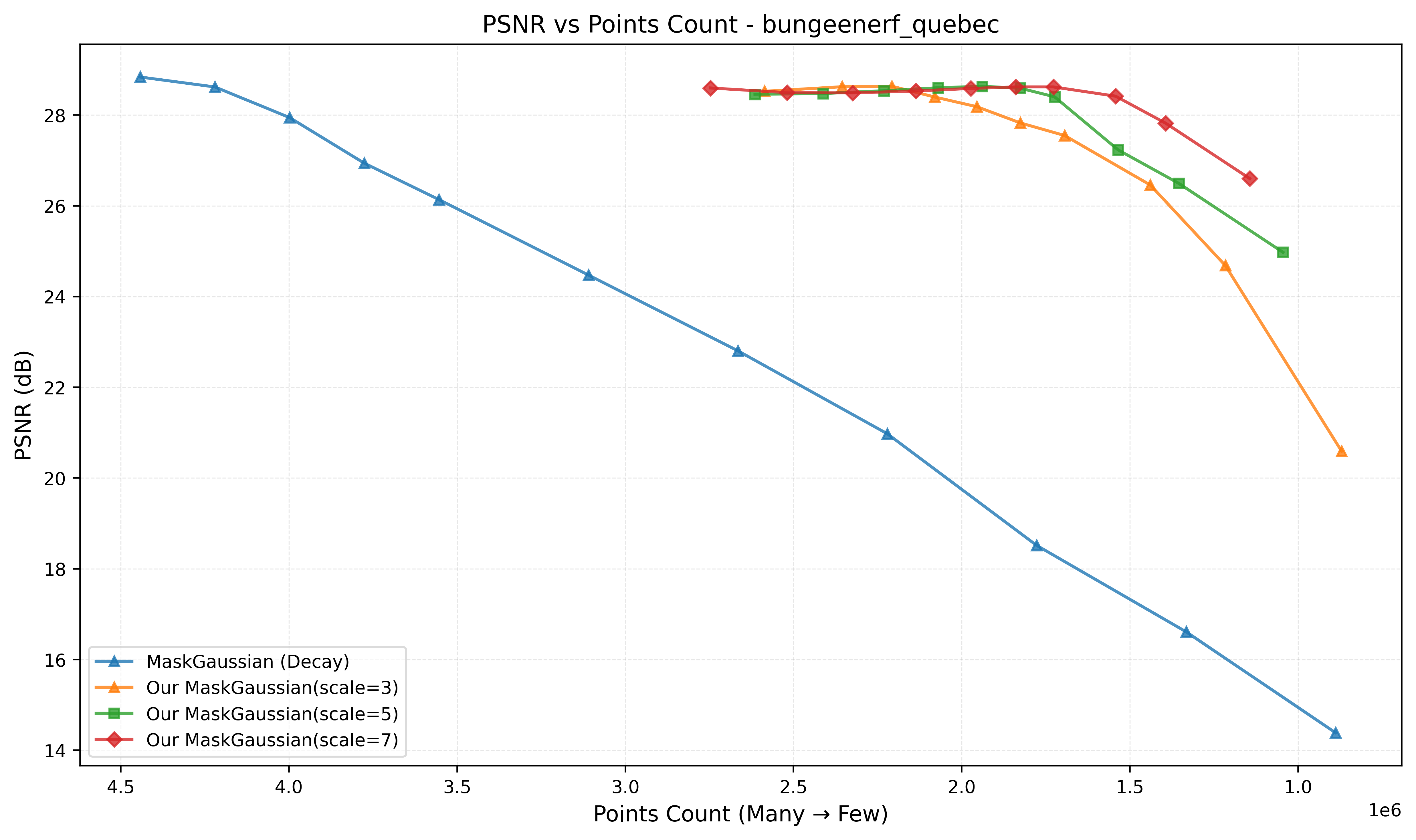}
\caption{MaskGaussian: PSNR vs. primitive count on the BungeeNeRF quebec dataset.}
\label{fig:curve_mask_quebec}
\end{figure}

\begin{figure}[h!]
\centering
\includegraphics[width=\textwidth]{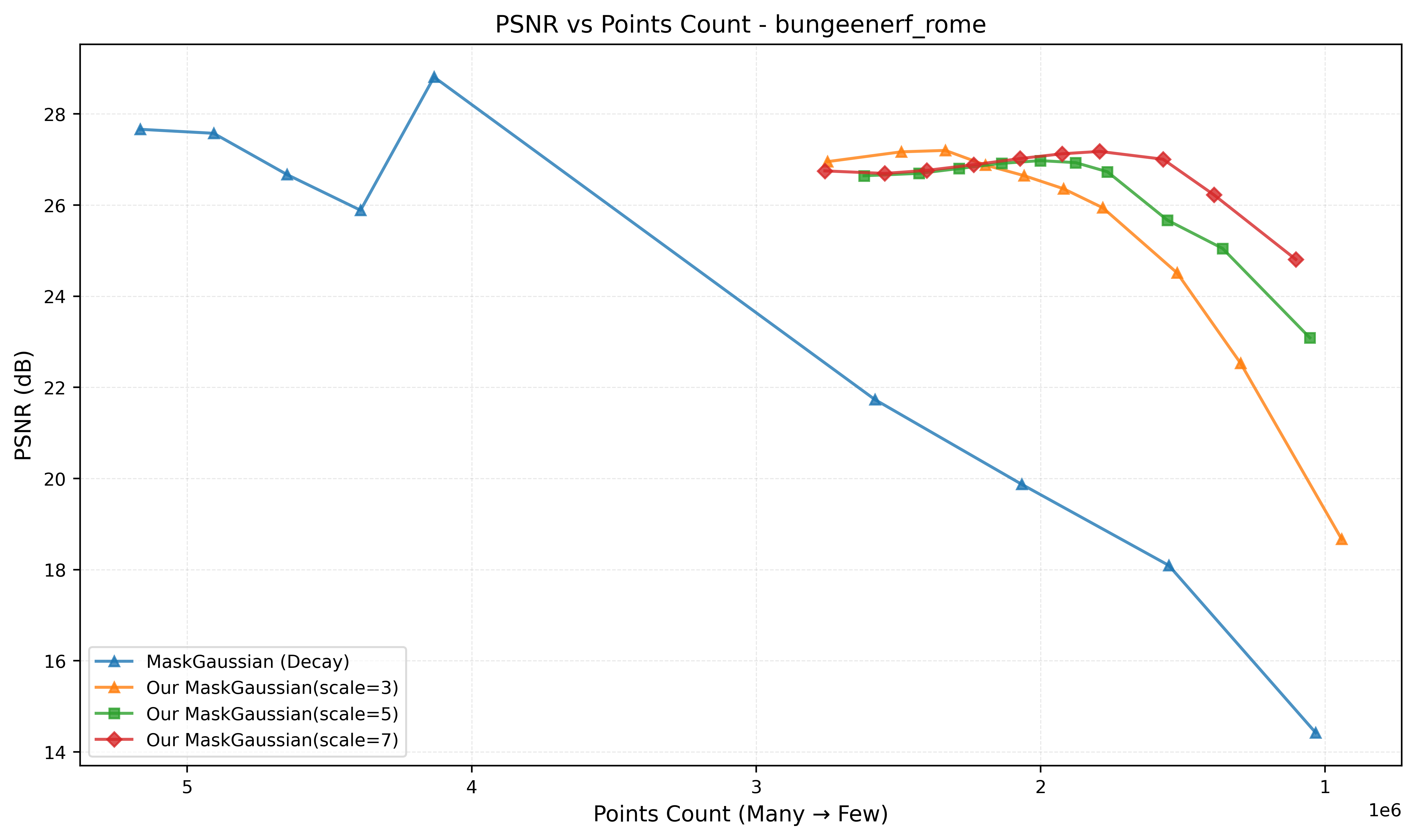}
\caption{MaskGaussian: PSNR vs. primitive count on the BungeeNeRF rome dataset.}
\label{fig:curve_mask_rome}
\end{figure}

\section{Use of LLMs}
During the preparation of this manuscript, we utilized the Large Language Model (LLM) Gemini 2.5 Pro\footnote{https://gemini.google.com/}, developed by Google. Its role was strictly limited to that of a writing assistant. Specifically, the model was employed for proofreading and copy-editing to improve the grammatical accuracy, clarity, and overall readability of the text. The LLM was not used for generating core research ideas, developing the methodology, conducting experiments, analyzing results, or drawing scientific conclusions. All intellectual contributions, including the concepts, experiments, and conclusions presented in this paper, are solely the work of the human authors.

\end{document}